\documentclass[%
 reprint,
superscriptaddress,
 amsmath,amssymb,
prd,
twocolumn
]{revtex4-2}

\usepackage{graphicx}
\usepackage{dcolumn}
\usepackage{bm}

\usepackage{hyperref}
\usepackage{mathrsfs}
\usepackage[caption=false]{subfig}
\usepackage[dvipsnames]{xcolor}
\definecolor{myred}{rgb}{0.89,0.259,0.204}

\usepackage{txfonts}
\newcommand{\comment}[1]{}

\begin{document}

\title{Dark magnetohydrodynamics: Black hole accretion in superradiant dark photon clouds }

\author{Shuo Xin}
\affiliation{SLAC National Accelerator Laboratory, Stanford University, Stanford, CA 94309, USA}
\affiliation{Physics Department, Stanford University, Stanford, CA 94305, USA}
\author{Elias R. Most}
\affiliation{TAPIR, Mailcode 350-17, California Institute of Technology, Pasadena, CA 91125, USA}
\affiliation{Walter Burke Institute for Theoretical Physics, California Institute of Technology, Pasadena, CA 91125, USA}

\begin{abstract}
  Black holes threaded by massive vector fields can be subject to a
superradiant instability, growing a cloud of massive vector particles
around it. In this work, we consider what happens if such a dark matter candidate field
mimicking a dark photon interacts with an accretion flow onto the black
hole. By including a kinetic mixing term with the standard model photon, we extend
the commonly used equations of general-relativistic magnetohydrodynamics to
a dark photon constituent. The coupling to the dark photon then appears as
an effective dynamo term together with a dark Lorentz force
acting on the accreting matter.  We numerically
study the interactions between the superradiant dark photon cloud and the inner
accretion flow by solving the coupled system in full numerical relativity.
By parameterically varying the mixing parameter between dark and standard
model sector, we provide a first investigation of how the accretion flow
could be modified. Depending on the coupling strength, our
solutions exhibit increased wind launching, as well as oscillation modes in
the disk.
\end{abstract}
\maketitle

\section{Introduction}

The study of physics near strong gravity sources has drawn increasing
attention. Gravitational wave detections of merging stellar mass compact
objects have provided new insights into the theory of relativity
\cite{Brito:2015oca,Berti:2015itd,PhysRevD.94.084002,fan2017speed,
cornish2017bounding,Isi:2017equ,Isi:2019aib,Cardoso:2019rvt}. 
With upcoming space-based detectors \cite{lisa_org}, as well as potential constraints on
the emission geometry of black holes \cite{konoplya2019shadow,gralla2019black,broderick2022photon},
we have excellent opportunities to test extensions of general relativity
\cite{kramer2006tests,yunes2013gravitational,Cornish:2017oic}. Apart from modified gravity, black holes (BH) also have the
potential to probe new states of (dark) matter \cite{Berti:2022wzk,Baryakhtar:2022hbu}.
Many different dark matter candidates have been brought forward including ultralight vector
bosons \cite{baumann2019probing,isi2019directed,bustillo2021gw190521}, axions \cite{svrcek2006axions,arvanitaki2010string}, scalar fields \cite{nusser2005structure} to just name a few.
Understanding whether and how they could modify the dynamics of gravitational wave emission 
and matter flows (environmental effects) around black holes is one of the key challenges for the
LISA mission \cite{Amaro-Seoane:2012aqc,eLISA:2013xep}.


One particular class of models are ultralight massive bosons, e.g. the QCD axion
\cite{weinberg1978new,svrcek2006axions, arvanitaki2010string} and dark photon
\cite{okun1982limits, feng2009hidden, cyr2013cosmology,bhoonah2019galactic}.
Adding to their appeal in this context, massive boson fields threading
black holes can result in superradiance \cite{PhysRevD.7.949,arvanitaki2010string}. 
For sufficiently high black hole spin, bosonic particles subject to
superradiance will undergo exponential growth outside the angular momentum barrier of
Kerr black holes, extracting rotational energy from the hole, and form a macroscopic
bound state. In other words, black holes may self-produce copious
amounts of dark matter through general-relativistic instabilities, making
them an ideal playground to investigate this kind of dark matter models.
Over the past years, a large body of literature has been
devoted toward understanding this phenomenon and its implications
analytically (e.g., 
\cite{PhysRevD.85.044043, richard2015superradiance,
arvanitaki2015discovering,arvanitaki2017black,
frolov2018massive,dolan2018instability,siemonsen2020gravitational}) and
through numerical simulations (e.g., \cite{witek2013superradiant,
zilhao2015nonlinear,east2017superradiant,east2017superradiant2,wang2022superradiance}).

At the same time, analytic computations of massive vector mode spectra have
matured considerably. Early calculations have provided approximate
characteristic frequencies and growth/decay rates 
\cite{konoplya2006massive,konoplya2007late,witek2011stability,rosa2012massive,pani2012black,pani2012perturbations,baryakhtar2017black,endlich2017modern}.
Recent work by Frolov, Krtous, Kubiznak and Santos (FKKS) on the
separability of the Teukolsky-like perturbation equation
\cite{frolov2018massive} has enabled subsequent analytic
calculations of quasi-normal modes to higher
accuracy\cite{dolan2018instability,siemonsen2020gravitational},
including potential gravitational wave signatures
\cite{zilhao2015nonlinear,east2017superradiant2}, see also Refs.
\cite{brito2017gravitational,Tsukada:2020lgt, Kalogera:2021bya,
Baryakhtar:2022hbu, jones2023methods} for proposed search strategies.

In addition to gravitational wave emission, a superradiant dark photon
cloud may also leave imprints through a potential coupling to the standard model. 
In many astrophysical environments, black holes are surrounded by gaseous accretion flows, powering strong outflows and electromagnetic
emission \cite{blandford1977electromagnetic,fishbone1976relativistic,
lee2000blandford, blandford2019relativistic}. Depending on the coupling strength, it is conceivable that 
interactions with the dark photon cloud through kinetic mixing terms in the Lagrangian
could modify astrophysical observables.
Recent work by Siemonsen et al. \cite{siemonsen2023dark} has studied the
dynamics of a force-free pair plasma modulated by interactions with such a  dark
photon cloud around an isolated stellar mass black hole formed in a merger, finding potential
X-ray signatures of dark matter interactions.

While force-free plasmas are applicable in the absence of external gas,
merging black holes in disk around active galactic nuclei (AGN), or even
the AGN itself will lead to the presence of matter and external magnetic
field. In this environment, rather then relying on magnetic field dynamics
entirely sourced by the dark photon cloud, the accretion flow will likely
be governed by the gas, with potential modifications through the dark
matter cloud. In the present work, we target precisely such an environment.

Due to the highly non-linear interplay of dynamical strong gravity, (dark) matter and electromagnetic dynamics, a full
investigation can only be carried out numerically. 
While accretion flows onto black hole mimickers have been investigated extensively
\cite{Mizuno:2018lxz,olivares2020tell,Chen:2021lvo}, works in dynamical
gravity have so far largely been limited to scalar field dynamics (see
\cite{siemonsen2023dark} for magnetospheric dynamics only). 

In this paper we numerically study accretion flows through
(self-gravitating) superradiant dark photon clouds in the dynamical
spacetime of a black hole. To this end, we derive a new set of evolution
equations that couple commonly used general-relativistic
magnetohydrodynamics (GRMHD) equations \cite{Shapiro2003} with those of a Proca field
\cite{zilhao2015nonlinear}. We provide a first assessment of the
resulting accretion flow structure and discuss potential implications of
the coupling between dark and standard model sector.

This paper is structured as follows. 
We briefly introduce the physics of superradiance and
GRMHD in Sec. \ref{p_physics} and explain our formulation and numerical
implementation in Sec. \ref{p_method}. Examples of black hole magnetospheric dynamics and accretion disks are shown in Sec.\ref{p_results} and we conclude in
Sec. \ref{p_conclusion}. Throughout this paper, we use geometrized units
($G=c=1$) and measure quantities by respective powers of the black hole mass
$M$. We adopt a four-dimensional metric $g_{\mu\nu}$ with signature $\left(-,+,+,+\right)$.

\section{Physical picture}

Superradiance is a prominent feature of massive bosonic fields around
spinning black holes. In black hole perturbation theory it is a family of
exponentially growing solutions of quasi-normal
modes (QNM). We here consider Newman-Penrose (NP) scalars $\psi$ constructed from
fields of spin weight $S$, or vector fields from $A_\mu = B_{\mu\nu}
\nabla^\nu \psi$ with polarisation tensor $B_{\mu\nu}$ of weight $S$
\cite{frolov2018massive,dolan2018instability,siemonsen2020gravitational},
in a Kerr black hole background with spin parameter $a$.
Separating the field into spherical, $S(\theta) e^{ i m\phi}$ and radial,
$R(r)$, parts
\begin{align}
\psi = e^{-i \omega t + m\phi} R(r) S(\theta)\,,
\end{align}
radial perturbations are governed by Teukolsky-like equations of the form
\cite{teukolsky1,teukolsky2,teukolsky3}
\begin{equation}
    \Delta^2\frac{d}{dr}\left(\frac{1}{\Delta}\frac{d
	R_{} }{dr}\right)-V(r)R_{}= 0,
\label{eq_Teukolsky}
\end{equation}
where $\Delta=r^2-2Mr+a^2$. Here $\omega$ is the effective mode frequency
of the oscillation.
For massless fields such as electromagnetic fields and gravitational perturbations,
$V(r)$ acts as an angular momentum barrier, which is only nontrivial around the black
hole ergosphere, allowing simple sinusoidal wave solutions near the horizon $r_+ =
M+\sqrt{M^2-a^2}$ and infinity (after proper transformations such as
Sasaki-Nakamura \cite{Sasaki:1981sx,Hughes:2000pf,Sasaki2003,Lo:2023fvv}). One prominent feature of this angular
momentum barrier is that the energy reflectivity is larger than 1 when the
frequency of the field perturbation is lower than the angular frequency of
the outer horizon $\Omega_+ = a/\left(r_+^2 + a^2\right) $. Heuristically
this is dual to the Penrose process of point particles, allowing the
possibility of extracting rotational energy from the black hole.

In analogy with bound orbits, if an additional outer barrier exists in the
potential $V(r)$, we have the possibility of forming bound states of dark
matter perturbations around the black hole. Near horizon features can modify $V(r)$ inside,
and depending on the reflectivity of the barrier near the horizon, this may
lead to a superradiant ergoregion instability (large reflectivity)
\cite{superrad_1,superrad_2,superrad_3} or wave echoes (small reflectivity)
\cite{nature17,Xin:2021zir,Maggio:2021uge,Conklin:2021cbc,Ma:2022xmp}. We may also alter the potential from the outside. When the
associated dark particle is of mass $m_\gamma = \hbar \mu$, additional features at large radii near $r \sim 1/m_\gamma$ could naturally arise\cite{arvanitaki2010string}. Such ultralight massive bosons
are also promising candidates for dark matter \cite{baryakhtar2022dark}.
Naively, we may expect particles propagating with real frequencies
$\omega_R \sim \mu$ to form a bound state and exponentially grow  due to
superradiance at a rate $\omega_I \propto (\Omega_+ - \mu)$, ultimately
stabilizing at ``floating'' orbits
\cite{press1972floating}. 

Perturbative calculations show that for the QNM of massive vector fields
characterized by overtone number $ n$, azimuthal ``magnetic'' number
$m$ and polarisation number $S$ (there are different definitions in
literature, we adopt the notation in works after FKKS
\cite{frolov2018massive,dolan2018instability,siemonsen2020gravitational}),
the frequency spectra, to leading order of $\mu$, is given by
\cite{PhysRevD.85.044043, richard2015superradiance,
arvanitaki2015discovering,baryakhtar2017black,arvanitaki2017black,
frolov2018massive,dolan2018instability,siemonsen2020gravitational}
\begin{equation}
    \omega_R \approx \mu \left( 1- \frac{(M \mu)^2}{2(|m|+ n+S+1)^2} \right),
\end{equation}
and growth rate
\begin{equation}
    \omega_I \approx \left( m \Omega_+ - \omega_R\right) (M\mu)^{4|m|+5+2S}.
\end{equation}
Strong superradiance is expected for (near) extremal spin, earlier numerical
simulations has also shown an $M \omega_I \simeq 10^{-4}$ for $a=0.99$
black hole \cite{PhysRevD.87.043513}. Also the rate is highly dependent on
the dark ``fine-structure'' constant
\cite{payne2022imprint,siemonsen2023dark} $\alpha ' = M \mu$. The growth is
most effective when the black hole mass $M$ and the dark photon mass $\mu$
satisfy
\begin{equation}
    \alpha' = 1.336 \left( \frac{M}{ 10^6 M_{\odot}} \right) \left(
    {\frac{10^{-16} {\rm eV}}{m_\gamma}} \right) \sim 0.1.
\end{equation}

After the superradiant growth, a stable ``cloud'' of dark photon is formed
around the black hole, mainly occupying the $ n=0,\, m=1,\, S=-1$
state. Oscillation of the field emit gravitational waves, making the field
decay with a power law over time, $t$, \cite{konoplya2007late,koyama2002slowly}
\begin{equation}
    \psi \sim t^{-p}.
\end{equation}
It can be shown that initially when the cloud grows $p = |m| + s +\frac
32$, which transitions to $p = -5/6$ in the saturated state \cite{konoplya2007late, koyama2002slowly}. 
This is also approximately observed in early
numerical simulations \cite{witek2013superradiant,zilhao2015nonlinear}.
Recent studies, especially after the separation of variables worked out by
FKKS \cite{frolov2018massive}, also show growing interest in higher
overtones, higher multiples, and subdominant polarisations. These
subdominant modes can occupy different spatial locations. Polarisations
$S=0,+1$ are peaked further away from the black hole while the dominant
modes with $S=-1$ are confined near the horizon. They also lead to beating
features in gravitational wave emission \cite{dolan2018instability,siemonsen2020gravitational}.

\begin{figure}
    \centering
    \includegraphics[width=9cm]{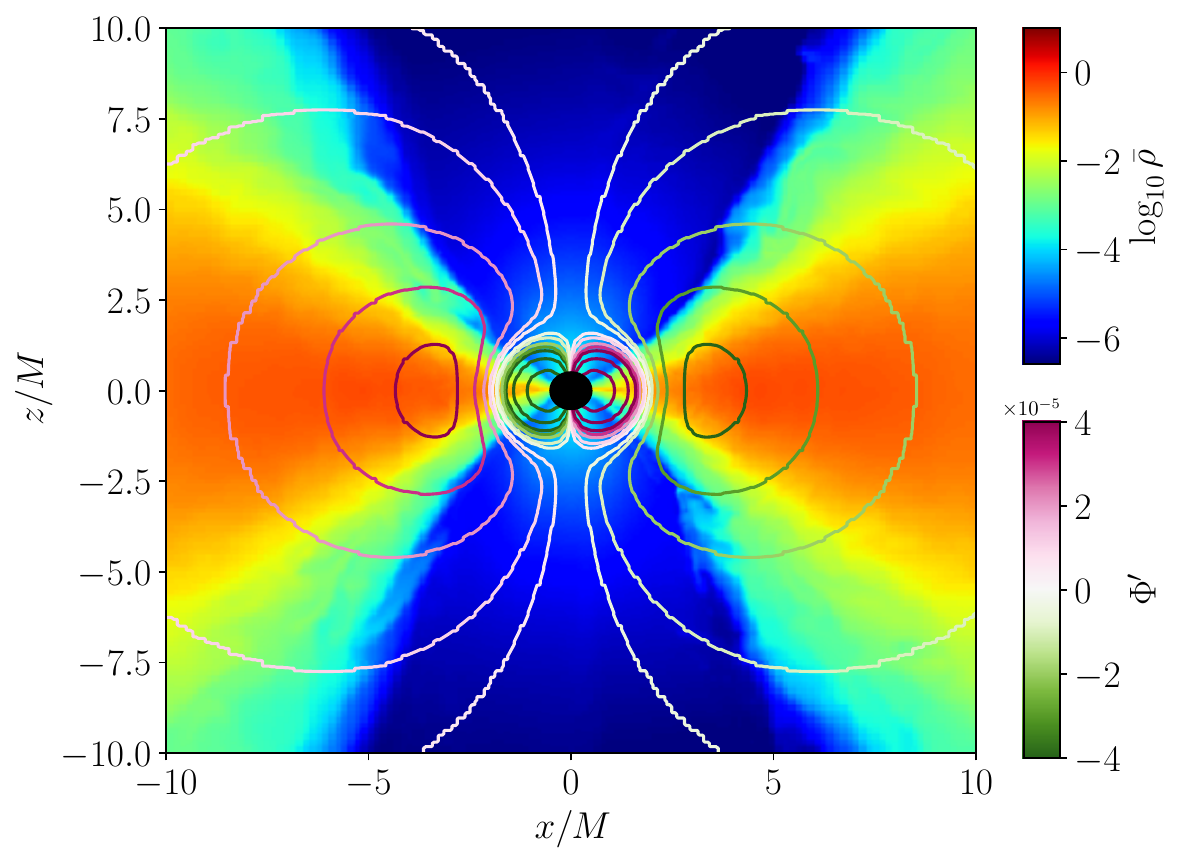}
    \caption{Schematic illustration of the overlap between the accretion disk
      and the dark photon cloud around a black hole (center). Shown in color are the rest-mass density,
      $\bar{\rho}$, of
      the accretion disk and contours of the Proca charge $\Phi ' = A'_\mu n^\mu $ in
      the dark photon cloud. Shown is an initial accretion snapshot in
      quasi-steady state after $t=2000M$, shortly before turning on the
      coupling of the dark photon field to the accretion flow, where $M$ is the black hole mass.  }
    \label{fig:disk_Aphi}
\end{figure}

Such macroscopic dark photon clouds can introduce rich physical effects by
coupling to the visible sector. Even if the electromagnetic environment
around the black hole is vacuum, the dark photon could potentially create
turbulent dynamics in the standard model matter, 
leading to dissipation and  emission \cite{siemonsen2023dark}.

In low energy effective theories that
are renormalizable, the relevant interaction is kinetic mixing through a
term $\epsilon \mu^2 A'_\mu A^\mu $ where $A_\mu$ is the vector potential
of visible electromagnetic fields and $A'_\mu$ of dark photon fields. As will be
detailed in Sec. \ref{p_method}, this leads to terms proportional to
$\epsilon {\alpha'}^2 {A'}^\mu $ in the effective Maxwell equations, normalized
by black hole mass $M$. Therefore the role of the superradiant cloud
depends on the relative scale of the dark $A'_\mu$ and the visible $A_\mu$
sectors.

The strength of the Proca fields is determined by the total energy endowed to
the cloud during the superradiant growth. This process has been intensively
explored both by numerical simulations \cite{east2017superradiant,
east2017superradiant2, witek2013superradiant, zilhao2015nonlinear} and
perturbative calculations
\cite{siemonsen2023dark,siemonsen2020gravitational}. This leads to a spin
down of the black hole from initial spins $a_i$ to final spins $a_f$, with rotational energy mainly
extracted by dark photon cloud. An estimate for the magnitude of the total
cloud mass $M'$ is\cite{siemonsen2023dark}
\begin{equation}
    \frac{M'}{M} \sim \alpha' \, (a_i - a_f).
\end{equation}
The radial distribution of this cloud is mainly confined inside a radius of
$r_0 \approx M/{\alpha'}^2$, as shown for the case of $\alpha ' =0.4$ in
Fig. \ref{fig:ProcaDecay}. Inside this radius the magnitude of dark
electromagnetic field is estimated by \cite{siemonsen2023dark}
\begin{equation}
    E' \sim (M')^{1/2} r_0^{-3/2},\quad B' \sim (M')^{1/2} r_0^{-5/2} \mu^{-1}.
\end{equation}

When coupled to the visible sector, this may introduce a modification of
the accretion flow structure near the black hole. 
As an order of magnitude estimate, for a strong ``fine-structure'' constant
$\alpha ' \sim 0.1$, the strength of this dark source corresponds to a
scale of visible magnetic fields of
\begin{equation}
\label{eq_B_estimate}
    B \sim 2.4 \left( \frac{M'}{10^{-2} M} \right) ^{1/2} \left({\frac{\alpha '}{0.1} }\right)^4 \left( \frac{\epsilon}{10^{-8}} \right) \left( \frac{10^6 M_\odot}{M} \right) \rm G,
\end{equation}
which is typical for magnetospheres and jet regions near supermassive black
holes in the parameter ranges we explore \cite{akiyama2021first}. Also note that in mass units,
such magnetic fields are of order $\sim 10^{-10} M^{-1}$ and the backreaction of it on the spacetime can be ignored, i.e. the magnetosphere and
accretion disk are not self-gravitating, while the dark sector is
macroscopic, comparable to black hole mass in extreme cases.


In simulations presented in this work, we explore parameters in ranges that
are both astrophysically common and allow for effective superradiance.  The
magnetic fields near black holes are of order up to $10^2$ Gauss and
electron density in the plasma is of order $10^{10} \rm cm^{-3}$. The black
hole spin is set to $a=0.9 M$ \cite{reynolds2021observational}. To allow for a significant amount of dark
photons we mostly explored the $\alpha '=0.4$ case when coupled to the
visible sector. The kinetic mixing (coupling) constant $\epsilon$ is of order $10^{-9}$ to
$10^{-7}$. The black hole mass only appears as an overall scale, and is understood to be between $10^6$ to $10^9$ solar mass to allow for
reasonable physical interpretation. Therefore, the dark photon mass range
we explore will be between $10^{-19}eV < m_\gamma < 10^{-16}eV$.  

 The energy and angular momentum of the superradiant dark photon cloud in
 the base state mainly resides in a toroidal region around the black hole,
 as shown both by perturbation theory calculations and numerical
 simulations
 \cite{east2017superradiant,east2017superradiant2,siemonsen2020gravitational,siemonsen2023dark,zilhao2015nonlinear}.
 This will allow for interesting interplay of it with accretion disks that
 astrophysical black holes often endow, as schematically illustrated in
 Fig. \ref{fig:disk_Aphi}. In our studies, we see large scale oscillation
 modes excited by the Proca field inside the disk and the launching of disk
 wind by mechanism beyond the traditional Blandford-Payne process due to
 the ``dark'' Lorentz force that we introduce in the next section.  

\label{p_physics}

\section{Methods}
\label{p_method}
\subsection{Dark magnetohydrodynamics in dynamical spacetimes}
\label{p_form}

Accretion disks are commonly observed around astrophysical black holes
\cite{begelman1984theory}. The relativistic dynamics of the plasma
and magnetic field in the strong gravity background is the central engine that powers accretion,
wind, and jets
\cite{blandford1982hydromagnetic,blandford1977electromagnetic,narayan1994advection,narayan1998advection,yuan2002jet,blandford1977electromagnetic,
fishbone1976relativistic, lee2000blandford, blandford2019relativistic}.
Accurate (GR-)MHD simulations are required for describing such astrophysical
settings. For our purpose, in addition, we also need to include the system
of equations for a massive Proca field, representing the dark photon. Due
to the superradiant nature of massive bosonic fields around spinning black
holes, the strength of the Proca field is macroscopic. The back-reaction of
them on the spacetime can, therfore, not be ignored. In the following, we
outline our approach towards the combined modeling of these effects.

The action that couples the electromagnetic vector potential $A^\mu$ and Proca field
$X^\mu $ (in mass basis) with gravity is
\cite{zilhao2015nonlinear,siemonsen2023dark,most2022electromagnetic}

\begin{equation}
\begin{aligned}
    \mathcal{S} = \int d^4 x \sqrt{-g} \Big( & \frac{R}{4} - \frac 14 W_{\mu\nu}W^{\mu\nu}- \frac 14 F_{\mu\nu}F^{\mu\nu} \\
    & - \frac{\mu^2}{2} X_\mu X^\mu + \frac{\epsilon}{2} F_{\mu\nu}W^{\mu\nu} + \mathcal{J}_\mu A^\mu  \Big),
\end{aligned}
\end{equation}
where we have defined the field strength tensors $F_{\mu\nu }=\nabla_{[\mu} A_{\nu]},
W_{\mu\nu} = \nabla_{[\mu} X_{\nu]} $.
Here, $R$ is the Ricci scalar and $\mathcal{J}^\mu$ the electric current. The coupling between the dark and visible sectors is expressed in
terms of a small coupling parameter $\epsilon$ \footnote{We use $\epsilon$ for the kinetic dark matter coupling and $\varepsilon$ for the Levi-Civita tensor.}.
Going to the interaction basis $
A'_\mu $ with $ X_\mu = A'_\mu  + \epsilon A_\mu $, $W_{\mu\nu} =
F'_{\mu\nu} + \epsilon F_{\mu\nu}$ where $F'_{\mu\nu }=\nabla_{[\mu}
A'_{\nu]}$: 

\begin{equation}
\begin{aligned}
\label{eq_Lag}
    \mathcal{S} = \int d^4 x \sqrt{-g} \Big( & \frac{R}{4} - \frac 14 F'_{\mu\nu}F'^{\mu\nu}- \frac 14 F_{\mu\nu}F^{\mu\nu} \\
    & - \frac{\mu^2}{2} A'_\mu A'^\mu -  \epsilon \mu^2 A'_\mu A^\mu  + \mathcal{J}_\mu A^\mu  \Big).
\end{aligned}
\end{equation}

The equations of motion for the Proca field $A'_\mu$ and the ordinary electromagnetic field then follow as,
\begin{align}
    \nabla_\nu F'^{\mu\nu} &=- \mu^2 A'^\mu + \epsilon \mu^2 A^\mu ,
    \label{eq_proca}\\
    \nabla_\nu F^{\mu\nu} &= \mathcal{J}^\mu - \epsilon \mu^2 A'^\mu .
\label{eq_EM}
\end{align}
As we see above, the dark photon effectively contributes as an additional
current $- \epsilon \mu^2 A'^\mu$ to the Maxwell equations.
The presence of this current has important consequences. As we can see in
the absence of a coupling, $\epsilon=0$, it immediately implies that the
dark photon field needs to be in Lorenz gauge, $\nabla_\mu A'^\mu=0$, which
in the presence of a coupling to the visible sector implies that the vector
potential $A^\mu$ needs to be in the same gauge, $\nabla_\mu A^\mu=0$.
Since $A'^\mu$ also describes the current, it further implies that the gauge
scalar $\Phi' \simeq  A'^0$ at the same time describes a dark charge. It
will therefore be convenient to visualize the dark photon field this way.\\

The main challenge is now to couple the dark and visible sectors in a way
compatible with the usual ideal MHD assumption, that the resistivity $\eta$
of the plasma vanishes, screening any electric field $e^\mu = u_\nu
F^{\mu\nu}$ in the fluid frame given by the plasma four-velocity $u^\mu$.
On a practical level, retaining the dark photon interaction current, will
naturally violate this assumption. However, following a similar approach
recently employed for the feedback of mean-field dynamo terms \cite{most2023impact}, we can perturbatively include the current in the induction equation
to leading order, i.e., $e^\mu \simeq \mathcal{O}\left(\epsilon
\right)$, without changing the main character of the evolution system. Such
a coupling is naturally appropriate if the impact of the dark photon on the
plasma is small, or more precisely if the coupling timescale is much longer
than the resistive timescale of the plasma.

Following \cite{most2023impact}, we formulate the total Ohm's
law of the system by borrowing from relativistic
two-fluid plasmas \cite{Most:2021uck}, 
\begin{align}
  \tau u^\nu \nabla_\nu \mathcal{J}^{\mu} = - \mathcal{J}^\mu - \epsilon \mu^2
  A'^\mu +  \eta^{-1} e^\mu \,,
  \label{eqn:Ohms}
\end{align}
where $\tau$ is the effective collisional timescale of the plasma. The
ideal MHD limit is obtained by assuming effective collisionality and
resistivity $\tau,\eta
\rightarrow 0$. In practice, in our simulations $\eta$ will be non-zero
effectively due to numerical resistivity from the discretization of the
equations. In this spirit, we now approach the perfectly conducting
limit, such that $\tau =0$, but the feedback of the dark sector will be
retained relative to the grid resistivity, i.e.,
\begin{equation}
\label{eq_ideal}
    e^\mu \approx \frac{\epsilon}{\sigma } \mu^2 A'^\mu,
\end{equation}
We use coupling constant rescaled by conductivity $\bar\epsilon =
\epsilon/\sigma$ to quantify this effect.

The equation of motion for the spacetime metric is the usual Einstein's equation
\begin{equation}
    R_{\mu\nu} - \frac 12 R g_{\mu\nu} = 8\pi  T_{\mu\nu}^{\rm total}\,,
\end{equation}
where $R_{\mu\nu}$ is the Ricci tensor and $T_{\mu\nu}^{\rm total}$ the total energy-momentum tensor.
The total energy momentum tensor contains contributions from the
electromagnetic fields, Proca field and the fluid. The specification of
$T_{\mu\nu}^{\rm total}$ in terms of fluid density, pressure and velocity
defines the property of the plasma we are considering.

We describe an accreting plasma around black holes as an ideal fluid with rest-mass density $\rho$ and 4-velocity $u_\mu$. The equations of motion for the fluid are given by continuity $ \nabla_\mu (\rho u^\mu) = 0 $ and conservation of total
energy-momentum $ \nabla_\nu T^{\mu\nu}_{\rm total}=0
$ \cite{shibata2005magnetohydrodynamics}, where the total energy-momentum
gets contribution from the electromagnetic field, Proca field and ideal
fluid:
\begin{align}
T^{\rm total}_{\mu\nu} &=  T^{\rm EM + fluid}_{\mu\nu} + T^{\rm Proca}_{\mu\nu}\,, \\
\label{eq_TmunuEMfluid}
     T^{\rm EM + fluid}_{\mu\nu} = &  F_{\mu \rho} F_\nu^\rho - \frac 14 g_{\mu\nu} F^{\alpha\beta} F_{\alpha\beta} + \rho h u^\mu u^\nu + p g^{\mu\nu} , \\
    T^{\rm Proca}_{\mu\nu} = &  F'_{\mu \rho} {F'}_\nu^\rho - \frac 14 g_{\mu\nu} {F'}^{\alpha\beta} F'_{\alpha\beta} \nonumber\\ 
     & \quad \quad + \mu^2 \left(  A'_\mu A'_\nu - \frac 12 g_{\mu\nu} {A'}^\alpha A'_\alpha \right),
\end{align}
where $p$ is pressure and $h$ is the specific enthalpy. 
We have also included the electromagnetic contribution in a combined tensor $T^{\rm EM + fluid}_{\mu\nu}$. 

\subsection{3+1 decomposition}
\label{p_3plus1}

In this work we adopt the 3+1 decomposition of the spacetime metric, $g_{\mu\nu}$, \cite{gourgoulhon2007}
\begin{align}
ds^2 = (-\alpha^2 + \beta_i \beta^i ) dt^2 + 2 \beta_i dx^i dt + \gamma_{ij} dx^i dx^j\,,
\end{align}
where $\alpha$ is the lapse function, $\beta^i$ the shift, and $\gamma_{ij}$ the induced metric on the hypersurface normal to $n_\mu = \left(-\alpha, 0 , 0, 0\right)$.
The metric dynamics in the 3+1 decomposition is governed by the ADM equations \cite{Shapiro2003,shibata2005magnetohydrodynamics} with total energy-momentum tensor given above:
{
\begin{equation}
\label{eq_ADM}
    \begin{aligned}
        \partial_t \gamma_{ij} = & - 2 \alpha K_{ij} + \mathcal{L}_\beta \gamma_{ij} \\
        \partial_t K_{ij} = & - D_i D_j \alpha + \alpha (R_{ij} - 2 K_{ik} K^k_j + K K_{ij} ) \\
        & + \mathcal{L}_\beta K_{ij} - 8 \pi \alpha \left( S^{\rm total}_{ij} - \frac 12 \gamma_{ij} (S^{\rm total}-\rho) \right)
    \end{aligned}
\end{equation}
where $D_i$ is covariant derivative with respect to spatial metric $\gamma_{ij}$, $\mathcal{L}$ denotes Lie derivative, $K_{ij}$ is extrinsic curvature and the evolution equation of spatial metric is the definition of $K_{ij}$. $S^{\rm total}=T^{\mu\nu}_{\rm total} n_\mu n_\nu$ and $S^{\rm total}_{\alpha\beta} = T^{\rm total}_{\mu\nu} \gamma^\mu_\alpha \gamma^\nu_\beta $ are normal and spatial projections of total energy-momentum tensor. 
We solve the ADM equations using the augmented Z4c system
\cite{PhysRevD.52.5428,PhysRevD.59.024007}.
}

We can further simplify the expression for the energy momentum tensor. Combining the relation between the
normal and comoving electric field
\begin{equation}
     E^\mu = n_\nu F^{\mu\nu}= - \varepsilon^{\mu\nu\alpha} v_\nu B_\alpha + \Gamma^{-1}(e^\mu + n^\mu e^\nu n_\nu),
\end{equation}
with the expression for the dark matter induced non-ideal electric field \eqref{eq_ideal}, we find
\begin{equation}
\label{eq_E_dark_current}
    E^\mu = - \varepsilon^{\mu\nu\alpha} v_\nu B_\alpha + \bar \epsilon \mu^2 \Gamma^{-1} \mathcal{A'}^\mu,
\end{equation}
where $\mathcal{A'}^\mu = {A'}^\mu +n^\mu {A'}^\nu n_\nu $ is the spatial
projection of Proca field. We have also introduced the Lorentz factor $\Gamma = -n_\mu u^\mu$.
Here, the first term in Eq. \eqref{eq_E_dark_current} is the usual ideal electric field, whereas the second term represents a dark component. This is akin to an $\alpha-$term in the induction equation \cite{most2023impact},albeit here the standard model magnetic field will be driven to form loops around the dark vector potential.\\

Using this expression for the electric field, we can now write an evolution equation \eqref{eq_EM} for the standard model magnetic vector potential in 3+1 form,
\begin{align}
    \partial_t \mathcal{A}_i  - \varepsilon_{ijk} (\alpha v^j - \beta^j) B^k + \partial_i(\alpha \Phi - \beta^j \mathcal{A}_j) &= - \alpha \bar\epsilon \mu^2 \Gamma^{-1} \mathcal{A}'_i \,,\\
    \partial_t (\sqrt \gamma\Phi) + \partial_j (  \sqrt{\gamma} \alpha \mathcal{A}^j - \sqrt{\gamma} \beta^j \Phi ) &= - \alpha \kappa \sqrt{\gamma} \Phi \,,
\end{align}
where we split $A_\mu$ into spatial component $\mathcal{A}_i$ and time component $\Phi = -n^\mu A_\mu$, $v^i$ is the 3-velocity of fluid related to 4-velocity by $u^\mu = \Gamma (n^\mu + v^\mu)$. The second equation is the Lorenz gauge condition, where we have added an extra damping term parameterized by $\kappa$ in order to maintain numerical stability \cite{etienne2011relativistic}.\\

We now turn to the problem of formulating evolution equation for the hydrodynamics sector.
Following \cite{Shapiro2003}, we can express the generic energy momentum tensor of resistive GRMHD as
\begin{equation}
\begin{aligned}
    T^{\mu\nu}_{\rm MHD} \equiv&\, T^{\rm EM + fluid}_{\mu\nu} = \rho h u^\mu u^\nu +  \frac 12 (E^2 + B^2 + 2 p) g^{\mu\nu} \\ 
    &- E^\mu E^\nu - B^\mu B^\nu  + ( n^\mu \varepsilon^{\nu \alpha \beta } + n^\nu \varepsilon^{\mu \alpha \beta } ) E_\alpha B_\beta .
\end{aligned}
\end{equation}
The Proca contributions lead to two modifications to ideal MHD: a) an additional nonideal current Eq. \ref{eq_ideal}, b) an external source term to the energy momentum conservation law. We give a brief summary in the following paragraphs, and refer to Appendices \ref{app_fluid} and \ref{p_c2p} for a comprehensive discussion.

While we impose joint conservation of matter and visible EM fields, the presence of a self-gravitating dark matter cloud causes a dark Lorentz force, $I^\mu$, to act on the matter sector, i.e., 
\begin{equation}
\label{eq_Tmunu_mod}
    \nabla_\nu T^{\mu\nu}_{\rm MHD}  = - \nabla_\nu T^{\mu\nu}_{\rm Proca} =  \epsilon \mu^2 F'^{\mu\nu}A_\nu =: I^\mu \, .
\end{equation}

Thus, performing the usual decomposition of the energy-momentum tensor (with subscript $_{\rm MHD}$ omitted hereafter), $ S = T^{\mu\nu} n_\mu n_\nu,  S_\alpha = - T_{\mu\nu} n^\mu \gamma^\nu_\alpha ,  S_{\alpha\beta} = T_{\mu\nu} \gamma^\mu_\alpha \gamma^\nu_\beta \, $, we have the conserved MHD energy and momentum densities
\begin{equation}
\begin{aligned}
    S = & \rho h \Gamma^2 - p + \frac{E^2 + B^2}{2}, \\
    S_i = & \rho h \Gamma^2 v_i + \epsilon_{ijk} E^j B^k.
\end{aligned}
\end{equation}
where the electric field $E^i$ now includes an additional nonideal contribution from Eq. \ref{eq_E_dark_current}. This requires a revised primitive recovery algorithm \cite{Noble:2005gf,kastaun2021robust}, which we provide in Apppendix \ref{p_c2p}.

The modified hydrodynamic equations for the conserved quantities abvove in 3+1 form are ({following the derivation detailed in Appendix \ref{app_fluid}}):
\begin{align}
\label{eq_fluidrho}
\partial_t (\sqrt{\gamma} \Gamma \rho  ) & + \partial_j \Big(\sqrt \gamma (\alpha v^j - \beta^j) \Gamma \rho\Big)=0 \, ,\\
\label{eq_fluidS}
    \partial_t \left( \sqrt\gamma S\right) & - \partial_i \left[ \sqrt\gamma (\beta^i S - \alpha S^i) \right]\nonumber \\
    & = \sqrt{\gamma} \left[ \alpha K^{ij} S_{ij} - S^i D_i \alpha + \alpha \mathcal{I}_\Phi \right] \, , \\
\label{eq_fluidSi}
    \partial_t  \left( \sqrt\gamma S_j \right) & + \partial_i \left[ \sqrt\gamma (\alpha S^i_j - \beta^i S_j) \right] \nonumber \\
    & = \sqrt\gamma \left[ S_i \partial_j \beta^i - S\partial_j \alpha + \frac 12 S^{ik}\partial_j \gamma_{ik} + \alpha \mathcal I_j \right] \, ,
\end{align}
with $\mathcal I_\Phi = - n^\mu I_\mu$ and $\mathcal{I}_\mu = I_\mu - n_\mu \mathcal{I}_\phi $ the normal and spatial components of the energy momentum source $I_\mu$. 

\begin{figure*}[htb]
    \centering
    \includegraphics[width = 18cm]{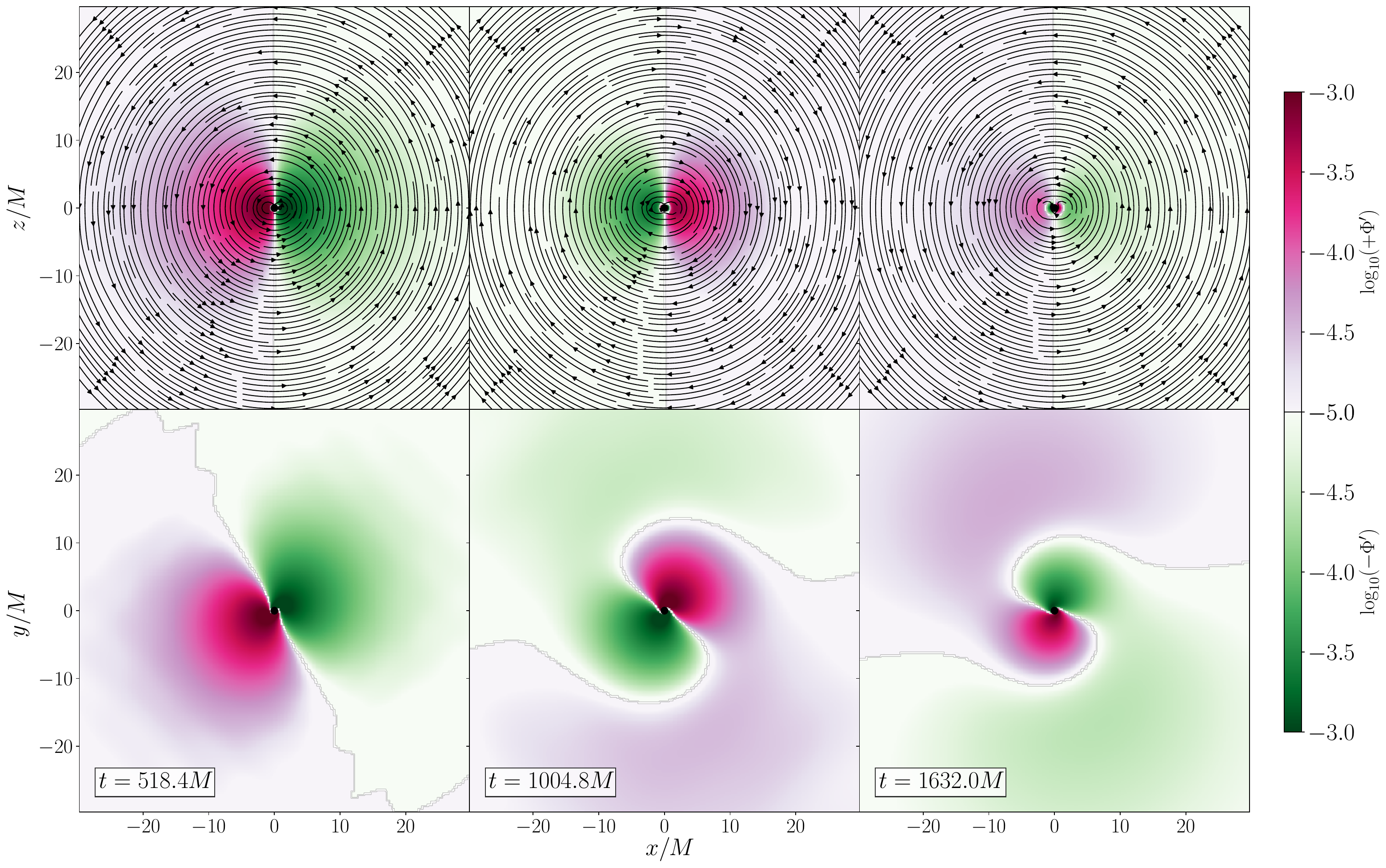}
    \includegraphics[width = 16.5cm]{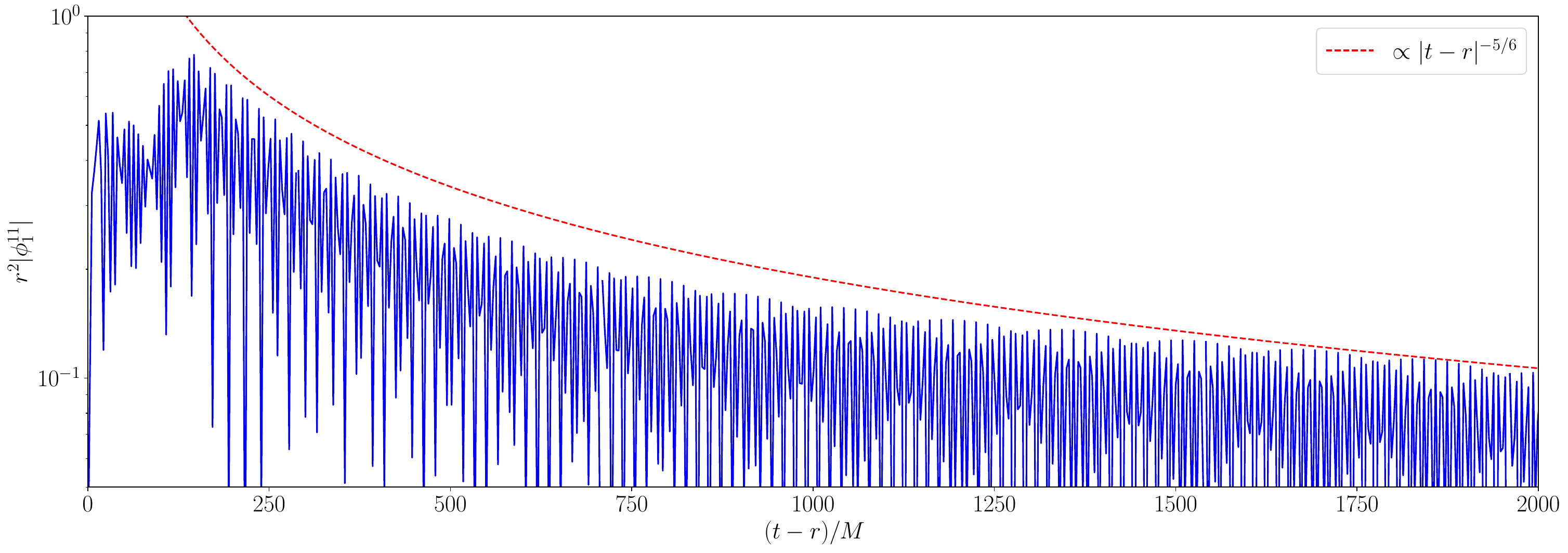} \hspace{1cm}
    \caption{
    Stable configuration (top) and power-law decay (bottom) of the dark photon cloud. Black solid arrows indicate the stream lines of the dark magnetic field. The strength of the Proca normal component $\Phi'$ is color coded in blue-red, magnitudes are measured in geometric units. The blue solid curve is the time evolution of the Proca Newman-Penrose scalar $\phi^{11}_1$ extracted at $r=100M$, where $M$ is the black hole mass. The power-law decay of the field is proportional to $\left|t-r\right|^{-5/6}$ and is shown with a red dashed curve. The initial condition shown is set according to Eq. \ref{eq:ICProca} and \ref{eq:GaussProca} with parameters $\hat C_{11} = 0.1$, $\hat r_0 = 6M$, $\hat \sigma = M$, $M \mu = 0.4$. \comment{Add $M$ to times, i.e., $t= \ldots M$}}
    \label{fig:ProcaDecay}
\end{figure*}
{ 
Expanding the dark electromagnetic field strength tensor $F'_{\mu\nu} = n_\mu E'_\nu -n_\nu E'_\mu + \varepsilon_{\mu\nu\alpha} B'^\alpha $, we have the source to MHD energy-momentum in terms of Proca $E',B'$ fields:
\begin{align}
    \mathcal{I}^\mu & =  \epsilon \rho'_e \left(E'^\mu + \varepsilon^{\mu\alpha\beta} v^D_\alpha B'_\beta\right)\,, \label{eq_dark_Lorentz} \\
    \mathcal{I}_\Phi &= \epsilon \rho'_e E'^i v^D_i\,,\label{eq_dark_heating}
\end{align}
where we have introduced the effective advection speed seen by the dark force $\left(v^D\right)^i = A^i / \Phi$ and the effective dark charge density $\rho'_e = \mu^2 \Phi $. This can be understood from the interaction term $\epsilon \mu^2 A'_\mu A^\mu$ in the Lagrangian Eq. \ref{eq_Lag}, where $\mu^2 A^\mu$ can be interpreted as the four-current to the dark sector.
For the Proca field equations we split the Proca field as $ A'_\mu = \mathcal{A}'_\mu + n_\mu \Phi'$ with normal components $\Phi' = -n^\mu A'_\mu$.

These additional terms in hydrodynamic equations can be intuitively understood: $\mathcal{I}_i$ in Eq. \ref{eq_dark_Lorentz} is the dark photon version of a Lorenz force, while the normal part $\mathcal{I}_\phi$ in Eq. \ref{eq_dark_heating} is ``dark'' Ohmic heating.
}
 Overall the equations are very similar to the ones outlined for the EM fields. Writing them in the form of Refs. \cite{witek2011stability,zilhao2015nonlinear} but adding additional interaction with the ordinary electromagnetic field, we have

\begin{align}
    (\partial_t -\mathcal{L}_\beta )\mathcal A'_i = & - \alpha E'_i - D_i(\alpha \Phi') \, , \\
    (\partial_t -\mathcal{L}_\beta ) \Phi' = & \alpha K \Phi' - D_i (\alpha \mathcal{A}'_i) \, , \\
    (\partial_t -\mathcal{L}_\beta ) E'^i = & \varepsilon^{ijk}D_j (\alpha B'_k) + \alpha K E'^i \nonumber \\
    & \quad +  \alpha \mu^2 \mathcal{A}'^i - \epsilon \alpha \mu^2 \mathcal{A}^i \, ,
\end{align}
where the last line evolution equation for dark electric field $ E'^\mu  = n_\nu F'^{\mu\nu}$, gets additional contribution from coupling with visible sector. Here, the visible EM potential $\mathcal{A}^i$ appears as an effective current to dark electric field.


\begin{figure}
    \centering
    \includegraphics[width=8.5cm]{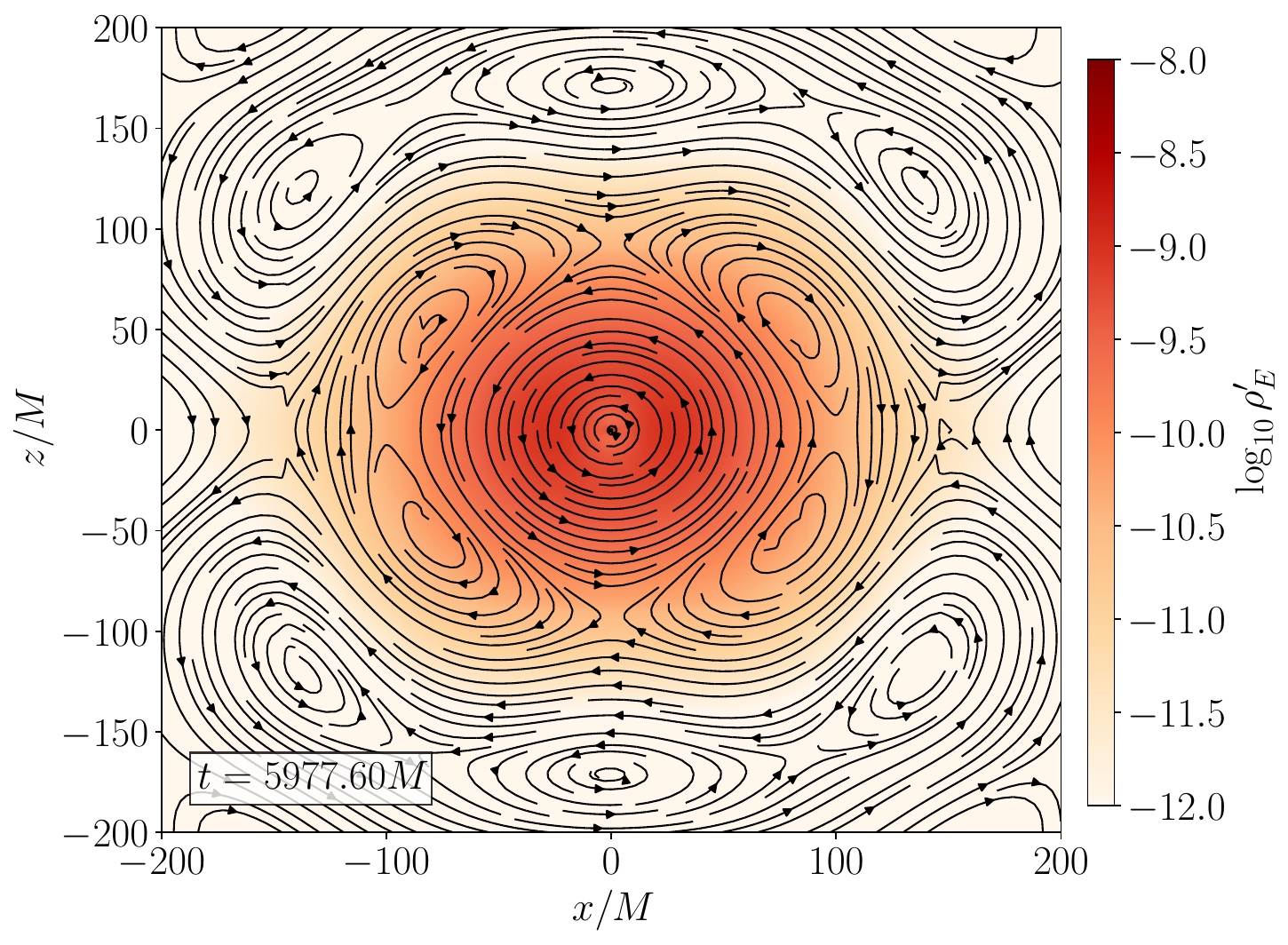}
    \caption{{Energy density $\rho'_E = n_\mu n_\nu T^{\mu\nu}_{\rm Proca}$ and magnetic field lines of the dark photon field. In addition to the dominant $S=-1$ polarisation, which peaks near the black hole and decays exponentially at large radii, the radial profile of energy density exhibit a subdominant mode peaked near $r\sim 40M$\cite{dolan2018instability}. Higher multipoles persist through late times. } }
    \label{fig:ProcarhoE}
\end{figure}


\begin{figure*}
    \centering
    \includegraphics[width = 6cm]{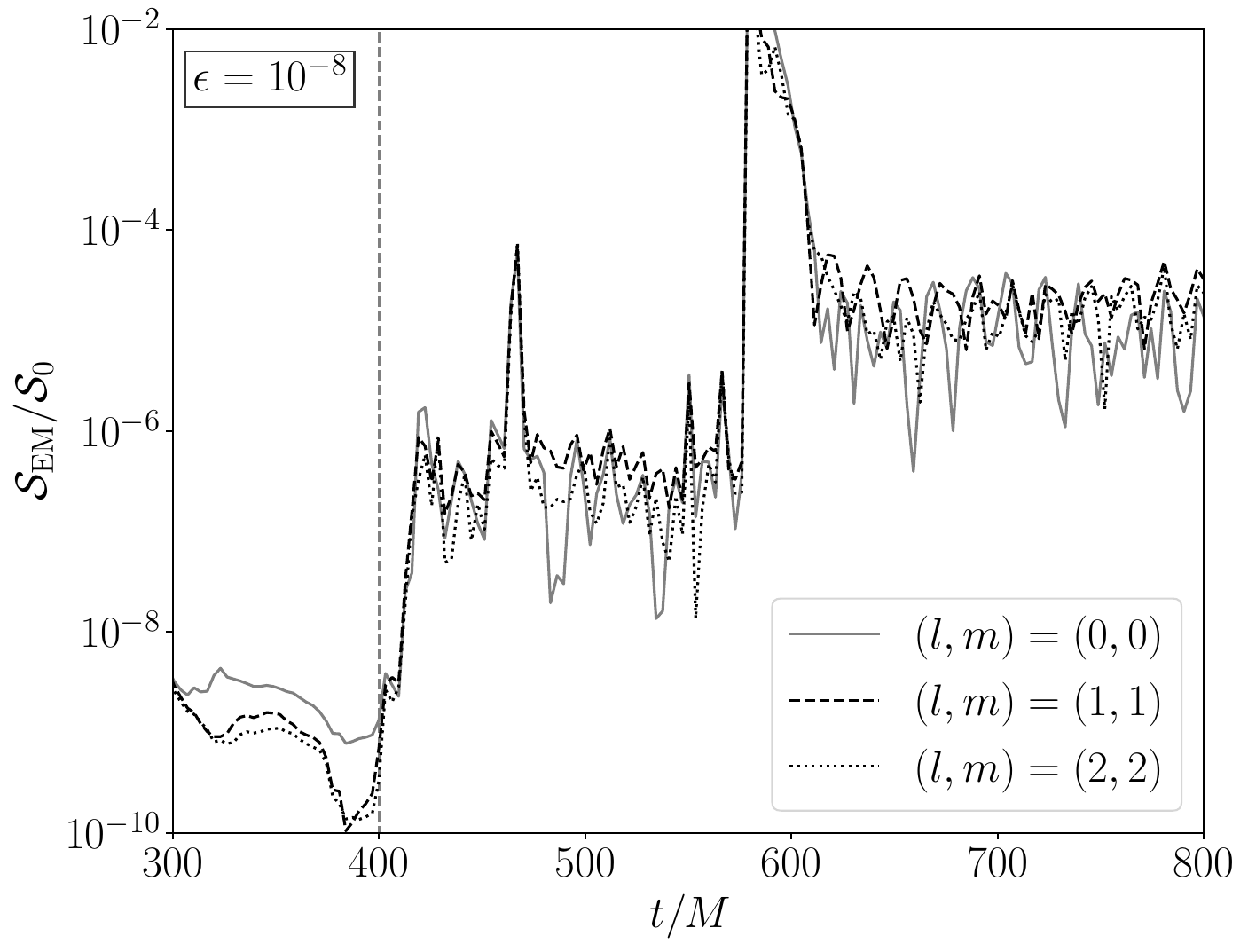}
    \includegraphics[width = 11cm]{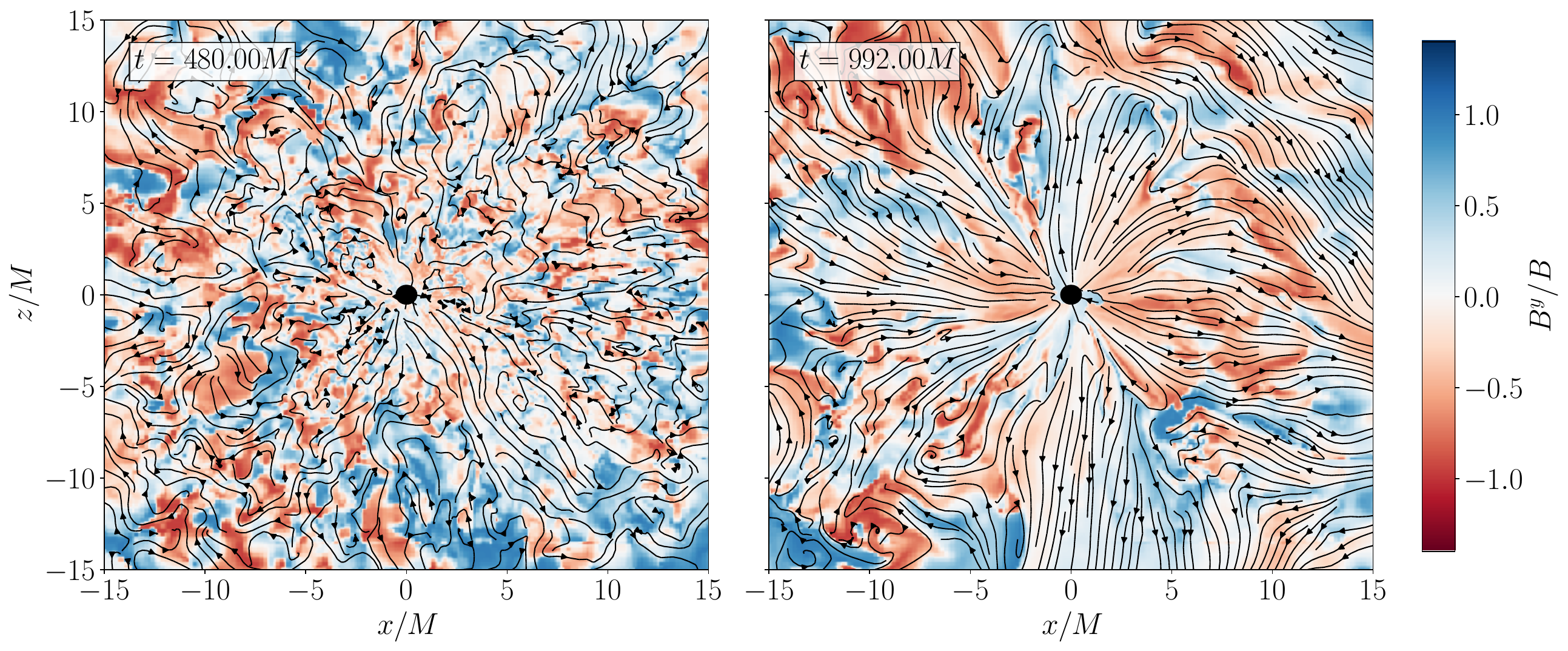}
    
    \includegraphics[width = 5.8cm]{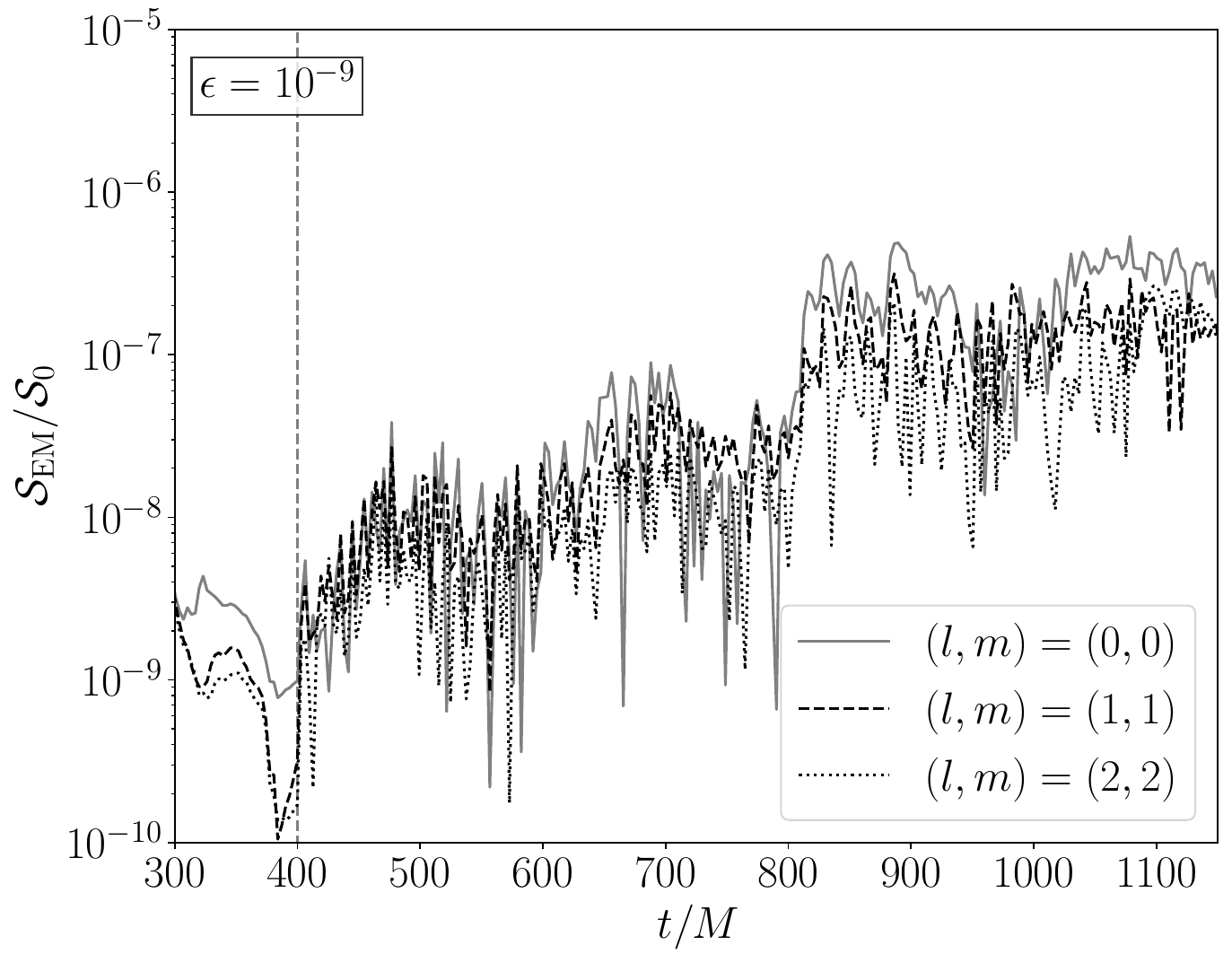}
    \hspace{0.2cm}
    \includegraphics[width = 11cm]{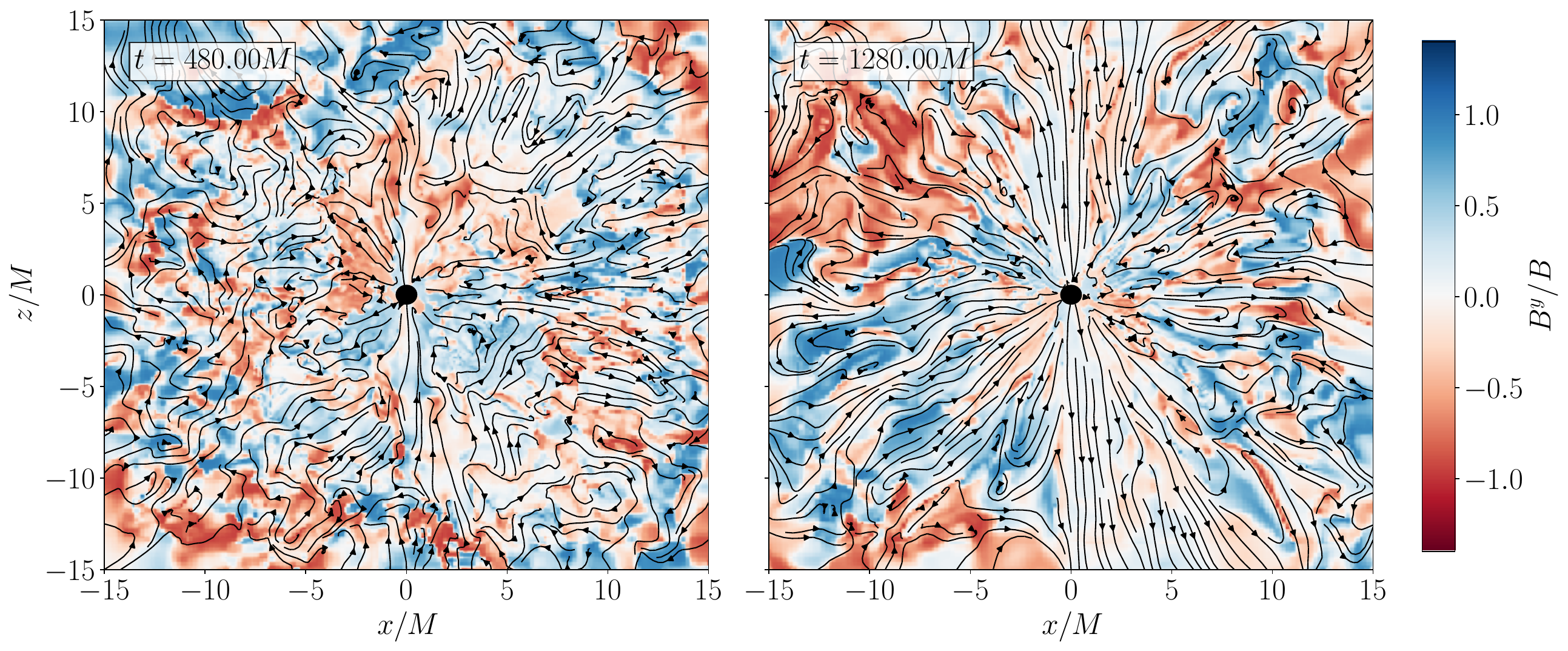}
    \caption{Electromagnetic energy (Poynting) flux sourced by the dark
      photon cloud (\textit{left}) and magnetic field configurations in the
      growth (\textit{middle}) and saturation stage (\textit{right}).
      Coupling constants are set to $\epsilon = 10^{-8}$ (top) and
      $\epsilon = 10^{-9}$ (bottom). As discussed in the main text, the
      Poynting scalar $\mathcal{S}_{\rm EM} = ( \vec E \times \vec B )
      \cdot \hat r $ is shown relative to $\mathcal{S}_0 = 10^{-16} M^{-2}$, where $M$ is the black hole mass.
      Shown is the strength of the out-of-plane magnetic field component, $B^y$, relative to the total field strength, $B$. Solid/dashed/dotted curves are
      $(0,0)$/$(1,1)$/$(2,2)$ modes of Poynting scalar extracted at radius
      $r=10M$. Vertical dash lines mark the time coupling is switched on
      ($t=400 M$). The streamlines on middle and left panels are visible
      magnetic field lines projected on the meridional plane. 
    }
    \label{fig:magnetic}
\end{figure*}


\subsection{Dark photon clouds around black holes}
\label{p_cloudsim}

To model a dark photon cloud around an isolated black hole, we adopt initial condition similar to the Gaussian initial condition in \cite{zilhao2015nonlinear}, where the normal component $\Phi'$ follows a Gaussian radial distribution and an angular distribution of $Y_{11}$ spherical harmonics:
\begin{equation}
\label{eq:ICProca}
	\Phi' = \psi^{-6} \hat C_{11} e^{-\frac{(r-\hat r_0)^2}{\hat \sigma^2}} Y_{11} (\theta, \phi),
\end{equation}
where $\psi^{12} = \det \{\gamma_{ij}\}$ is the usual coefficient in conformal decomposition of spacetime evolution equations, and $\hat C_{11},\hat r_0,\hat \sigma$ are the strength, center and width of the Gaussian distribution. The initial dark electric field $E'$ is set by solving a Poisson equation for the dark electric field:
\begin{equation}
\label{eq:GaussProca}
	D_i E'^i + \mu^2 \Phi' = 0,
\end{equation}
with the boundary condition $E'^i\rightarrow 0$ at $r\rightarrow \infty$. 

{
The characteristic time scale is set by the frequency of Proca field oscillation $1/\omega_R \sim 1/\mu \sim 2.5 M$. 
With initial setup $\hat C_{11} = 0.1$, $\hat r_0 = 6M$, $\hat \sigma = M$, $M \mu = 0.4$, the Proca field settles to a stable configuration after $t \sim 500 M$ of evolution time, as shown in Fig. \ref{fig:ProcaDecay}. The field normal component $\Phi '$ and dark magnetic field $\vec B'$ within $r<20M$ approximately follow $\ell=m=1$ spherical harmonics ($\ell,m$ understood as the usual angular and azimuthal numbers for spherical harmonics). This agrees with the perturbation calculation, where the dominant mode confined near black hole is $ n = 0, m=1,S=-1$ mode (the overtone, azimuthal, polarisation numbers for FKKS formulatiuon \cite{frolov2018massive,dolan2018instability,siemonsen2020gravitational}), whose angular distribution overlaps mostly with $Y_{11}$ spherical harmonics.
}

After the dark photon cloud is formed, it decays outward through emission of (gravitational and dark electromagnetic) radiation. We will find it useful to analyze this decay in terms of Newman-Penrose (NP) formalism. We set up the null tetrad ($k^\mu,l^\mu,m^\mu,\bar m^\mu$) by
\begin{equation}
\begin{aligned}
	k^\mu & = \frac{1}{\sqrt 2} (\hat n^\mu - \hat r^\mu),\\
	l^\mu & = \frac{1}{\sqrt 2} (\hat n^\mu + \hat r^\mu),\\
	m^\mu & = \frac{1}{\sqrt 2} (\hat \theta^\mu + i \hat \phi^\mu),
\end{aligned}
\end{equation}
where $\hat n$ is the unit vector in the normal $n^\mu$ direction, and ($\hat r, \hat \theta, \hat \phi$) is a basis of orthonormal spatial vectors which asymptotically behave as radial, polar and azimuthal vectors. The NP scalar of spin weight -1 for Proca field is (not to confuse with the notation for time component of electromagnetic field)
\begin{equation}
	\phi_1 = \frac 12 F'_{\mu\nu} (l^\mu k^\nu + \bar m ^\mu m^\nu).
\end{equation}

Different multipoles of $\Phi_1$ can be extracted by expansion
\begin{equation}
	\phi_1 = \sum_{\ell,m} \phi_1^{\ell m} Y_{\ell m}.
\end{equation}
In our case, the dominant multipole is $(1,1)$ mode $\phi^{11}_1$. The $(1,1)$ mode of Proca NP scalar of spin weight 0 is shown in the lower panel of Fig. \ref{fig:ProcaDecay}. It follows a $- 5/6$ power law decay at late times, as expected from perturbative calculations. Similar power law decays have also been observed by Ref. \cite{zilhao2015nonlinear}.

Higher order modes are present in our setup of the Proca field, which will play an important role in interaction with the visible sector. The dominant $ n=0,m=1,S=-1$ mode is confined near the horizon and exponentially decays with a characteristic radius $r\sim M/\mu^2$. This agrees with our simulation within around $r<10M$ for $\mu M = 0.4$ as shown in upper panel of Fig. \ref{fig:ProcarhoE}, as well as results in early simulations \cite{east2017superradiant,east2017superradiant2}. Perturbation calculations \cite{dolan2018instability} show subdominant modes of $S=+1$ polarisation are centered further away from the black hole, and will be important in its interaction with accretion disks and winds at large radii. Our simulations have physical extent of $x=\pm 1024 M$ and include the subdominant modes.

\subsection{Numerical implementation}
\label{p_sim}

Having outlined the mathematical description of the coupled Einstein-Proca-GRMHD (dark magnetohydrodynamics) system we are solving, we will now describe the numerical framework and setup used to carry out the simulations presented in subsequent Sections.\\

Our simulations make use of various codes based on, or provided by the \texttt{Einstein Toolkit} \cite{EinsteinToolkit}.
The Proca sector is solved using the publicly available suite of \texttt{Canuda} codes \cite{zilhao2015nonlinear,witek2021canuda,herdeiro2021imitation}, suitably modified for coupling them with a GRMHD evolution. In particular, we discretize the system using fourth-order finite differencing with Kreiss-Oliger dissipation. The use of unlimited finite-difference approaches is appropriate since the turbulent dynamics of the matter does not appreciably backreact on the dark photon cloud itself.
 The GRMHD equations are solved using the  \verb|Frankfurt/IllinoisGRMHD| (\verb|FIL|) code \cite{Most:2019kfe,Etienne:2015cea}.
 \texttt{FIL} provides a fourth-order accurate discretization of the GRMHD equation in vector potential form \cite{PhysRevD.85.024013}. On the technical side, the equations are solved using a modified ECHO scheme \cite{del2007echo}, with WENO-Z reconstruction and improved primitive recovery \cite{kastaun2021robust}, as used in recent GRMHD studies performed with \texttt{FIL} \cite{Most:2021ytn,Chabanov:2022twz,Most:2023sft,Most:2023sme,Most:2024qgc}.
 Due to the coupling with the Proca field, we needed to modify the primitive recovery method of Ref. \cite{kastaun2021robust}. A detailed description is provided in Appendix \ref{p_c2p}.
 
The spacetime is evolved using the \texttt{Antelope} module in \texttt{FIL} in the Z4c formulation\cite{Bernuzzi:2009ex,Hilditch:2012fp}.
We adopt moving puncture gauges \cite{alcubierre2005dynamical}, and discretize the equations using a fourth-order accurate finite-difference scheme \cite{zlochower2005accurate}.
  
The simulations were performed using the  \verb|EinsteinToolkit| infrastructure \cite{loffler2012einstein} with adaptive mesh-refinement provided by the \verb|Carpet| code \cite{Schnetter:2003rb}. 

In detail, we adopt the following grid setup:
Our simulation uses a 3D Cartesian grid with 9 static mesh refinement levels. The physical domain extends to $-1024 < x_i/M < 1024$ in all directions. The resolution doubles at each refinement level with the finest grid spacing being $\Delta x = 0.04M$. The initial spin parameters are set to $a=0.9$ for a highly spinning spacetime background. The initial spacetime is set by solving the puncture equation in the presence of a Proca field \cite{zilhao2015nonlinear,brandt1997simple}, using the \texttt{Canuda} code \cite{Canuda_2020_3565475}. 
  

\section{Results}
\label{p_results}

With our numerical implementation of the dark GRMHD system, interactions
between the dark photon cloud and various astrophysical environments can be
simulated. As a first test and application, we focus on black hole
accretion in two different regimes. First, we consider a regime where the
black hole is embedded in a low density gaseous environment, which is dominated by
the dark field, $A_\mu \ll \epsilon {A'}_\mu $ (Sec.
\ref{p_magnetosphere}).  Second, we consider an advection dominated
accretion flow \cite{Narayan:1994xi,Narayan:2008bv}, where the dark photon
field is subdominant to the visible magnetic field governing the accretion
flow $A_\mu \geq \epsilon A'_\mu $ (Sec. \ref{p_disk}).\\

In all of the flows we consider, we assume a mass hierarchy, in which the
fluid is not self-gravitating, the fluid motion does not affect the decay
and dynamics of the dark photon cloud.\\ In this way, the dark photon
cloud has a mass scale relative to the black hole mass, and the fluid only
acquires a mass scale through the mixing constant we choose.

Because the fluid quantities are much smaller than the gravitational scale
set by black hole mass $M$, for numerical accuracy, we rescaled the
magnetic fields and mass densities by small units of $B_0 = 10^{-8} M^{-1}$
and $\rho_0 = 10^{-16} M^{-2} $. The coupling constants we explored are
around $10^{-9} < \epsilon < 10^{-7}$ and the stable configurations of dark photon
cloud has $A'_\mu \sim 10^{-3}$. The frequency of the dark photon cloud is of order $ \omega_R \sim \mu = 0.4M^{-1} $. As estimated by Eq. \ref{eq_B_estimate}, influence of such dark photon field corresponds to a visible magnetic field at order $10^{-11}M^{-1} \sim 10^{-9}M^{-1}$. In the simulations presented in this paper, we explore two scenarios, the black hole with low density, low magnetization gas at order $B\sim 10^{-5}B_0$ where the dark photon cloud dominates the interaction, and the accretion disk with magnetic field in jet region of order $B\sim B_0$ where the dark photon cloud appears as a perturbation to MHD quantities.


\subsection{Black holes in low density nearly force-free environments}
\label{p_magnetosphere}

As a first step, we consider the evolution of a black hole in a low density force-free gas. Taking the limit of a dominant dark photon cloud, we can consider the scenario where the dark photon cloud will drive the generation of visible magnetic field and dynamics of the plasma via the dark Lorentz force and dynamo terms. Such a scenario has been recently considered by Ref. \cite{siemonsen2023dark} using dark force-free electrodynamics. In the limit of high magnetization $\sigma= b^2/\rho \gg 1$ our simulations are able to reproduce this limit, though in general subtle differences exist, including the potential for shock formation \cite{Most:2024qgc}.

\begin{figure}
    \centering
    \includegraphics[width = 8.2cm]{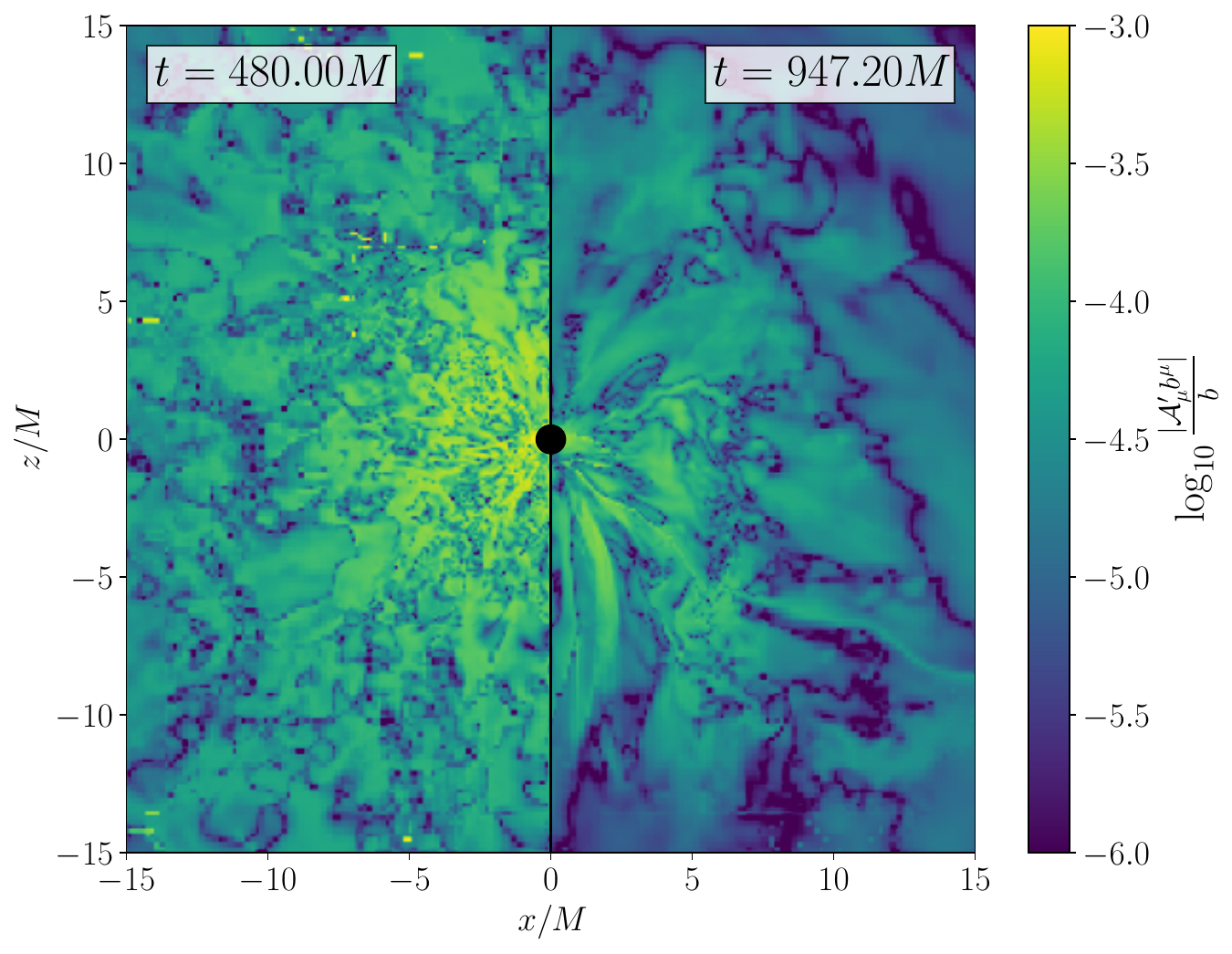}
    \caption{Effective $\alpha$ dynamo-term in the Ohms law for the
      force-free like isolated BH simulations. Shown are early and late
      times for the $\epsilon = 10^{-8}$ simulation shown in Fig.
      \ref{fig:magnetic}. }
    \label{fig:FFBH_alpha_sigma}
\end{figure}


The initial condition is motivated by a Wald scenario commonly studied for pair plasmas \cite{Parfrey:2018dnc,Komissarov:2021vks}. We initialize the system with vertical magnetic field, $B_z \ll \epsilon \Phi'$. If sufficiently magnetized, such that the system reaches a nearly force-free state, such a solution on its own will form a transient jet-like outflow by accreting vertical flux on the black hole. 
Instead, here we consider a case of low magnetization, $\sigma = 0.01$, for which accretion of plasma will largely be hydrodynamical. That is, without additional amplification of magnetic fields, which needs to be provided by an external energy reservoir, the system would not (on the timescales we consider here) become magnetically dominated or launch any outflows. As we will now show, energy conversion of the superradiant cloud into the magnetic field is able to drastically change this picture, as initially demonstrated by Ref. \cite{siemonsen2023dark}.\\

\begin{figure}[tb!]
    \centering
    \includegraphics[width = 8cm]{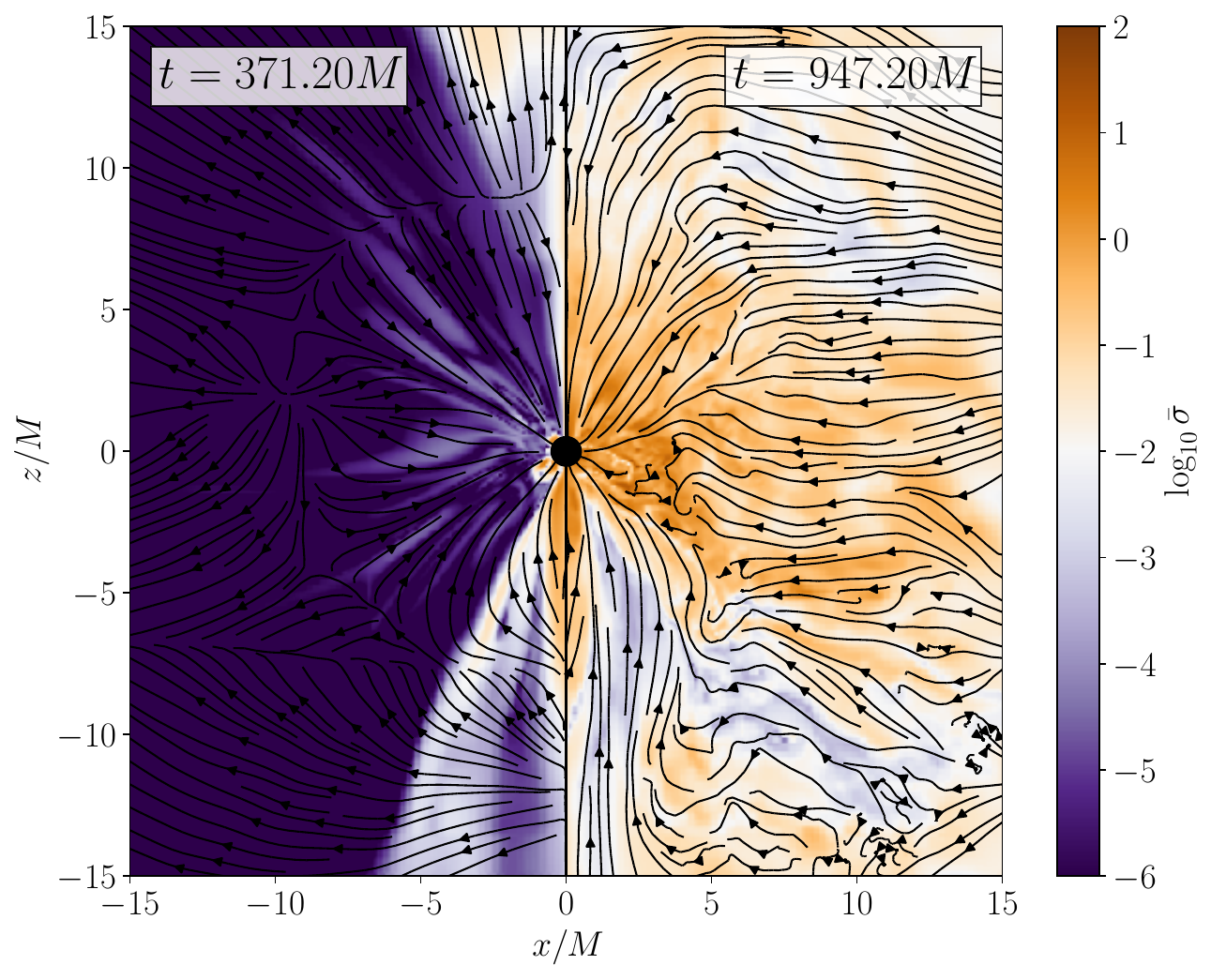}
    \caption{Magnetization parameter $\sigma = b^2/\rho$ for the strong coupling case $\epsilon=10^{-8}$.
    Streamlines include the velocity structure of the gas flow.
    The simulation is the isolated BH solution shown also in Fig. \ref{fig:magnetic}.}
    \label{fig:sigma_bh_ffe}
\end{figure}

After turning on the coupling, the dark photon field will interact with the magnetic field around the black hole and begin to amplify the field. We can see that initially at time $\Delta t=80 M$ after the coupling was turned on the field begins to get turbulently amplified (see Fig. \ref{fig:magnetic}). A similar observation was also made by Ref. \cite{siemonsen2023dark}. At later times, we can see that the magnetic field rearranges itself into a Wald like structure, around the black hole with large scale flux tubes emerging. This ordered field structure seems to settle into a constant outflow state as indicated by the Poynting flux measured close to the hole (Fig. \ref{fig:magnetic}). Overall, the dark photon cloud itself does not imprint its oscillation modes onto the Poynting flux, as multiple modes are present and of comparable strength. The coupling strength of the dark photon to the visible photon clearly impacts the amount of larger structure formation, with more large scale structures being present for the stronger coupling case ($\epsilon=10^{-8}$).
We can best quantify this effect by consider the dark contribution to the induction equation, which in the near ideal MHD regime we consider governs the evolution of the magnetic field. We find that, expressed as an $\alpha-$term, $A'^\mu b_\mu/b^2$, the system features substantial and strong contributions at early times. These drive turbulent dynamo amplification of the magnetic field in the vicinity of the black hole, which switches off at late times, with the magnetic field lines being largely perpendicular to the dark vector field $A'^\mu$. In addition, we quantify this saturated state also in terms of the magnetization $\sigma$, as shown in Fig. \ref{fig:sigma_bh_ffe}. Here, we can see that at late times the close vicinity of the black hole is in a nearly force-free state $\sigma > 1$. Velocity streamlines indicate that increased accretion onto the black hole.

\subsection{Black hole accretion}
\label{p_disk}

We now focus on the case of an accretion disk around a spinning black hole
commonly used to model accretion around compact objects \cite{Davis:2020wea}. 
We initialize the dark photon field as before, but this time
add a constant angular momentum torus with a vertical net magnetic fields
of magnitude $B_z \gg \epsilon \Phi'$.  In this case, the flow structure
will only be perturbatively affected by the Proca field, different from the
case considered in the previous section. We expect such a scenario to be
more realistic given that dark matter should only couple weakly to
accretion flows.

\begin{figure}[tb!]
    \centering
    \includegraphics[width = 0.5\textwidth]{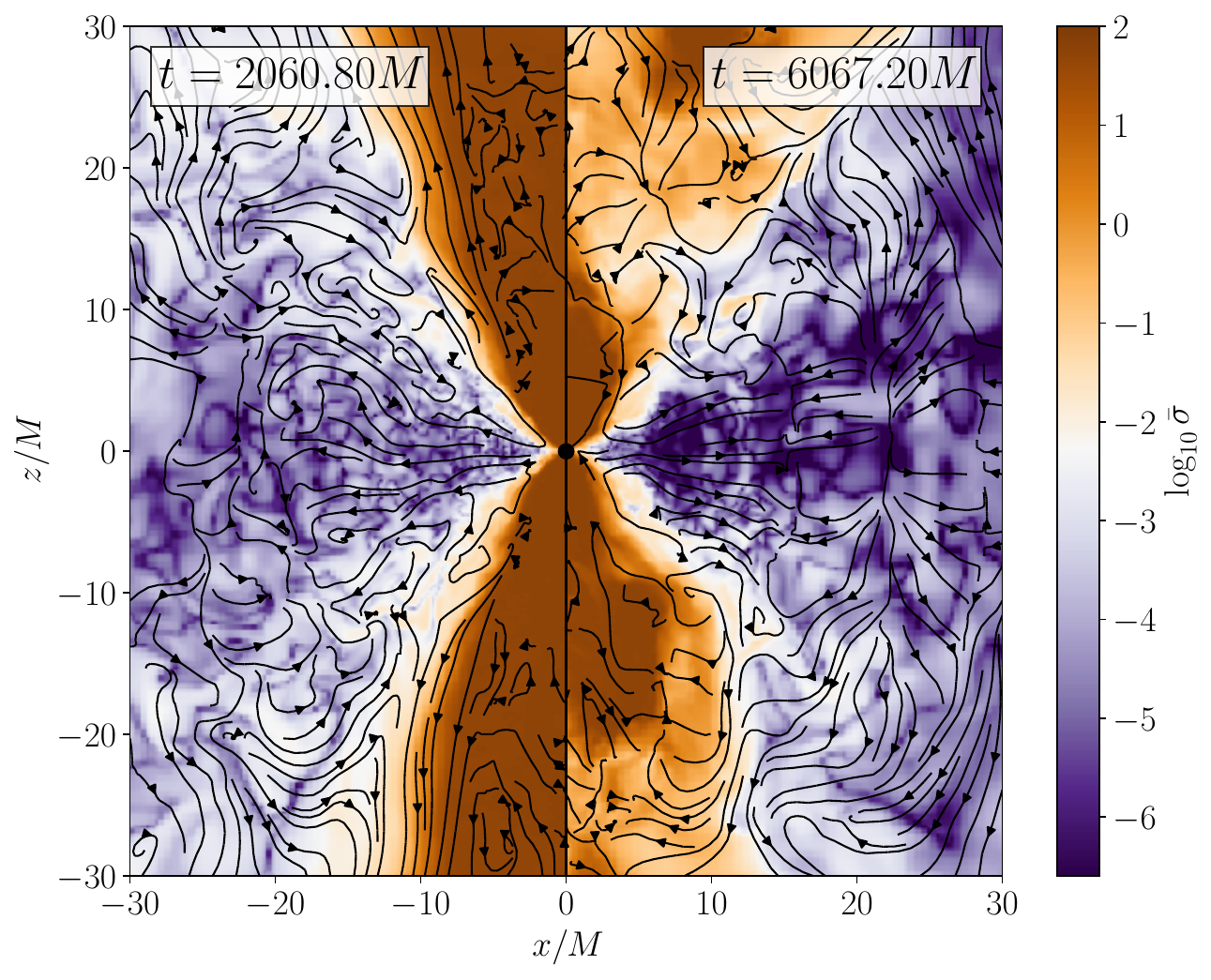}
    \caption{ Fluid velocity streamlines and distribution of magnetization $\sigma = b^2 / \rho $ for the accreting black hole scenario. The time, $t$, is shown relative to the start time of the simulation, where $M$ is the black hole mass. We activate the dark photon coupling once a quasi-steady state is reached corresponding roughly to the time shown in the left panel.}
    \label{fig:sigma_bh_acc}
\end{figure}

To obtain a quasi-equilibrium configuration of an accretion disk, the hydrodynamic quantities need to be correctly set in the initial condition to have a stable circulation of fluid around black hole. Due to the non-negligible self-gravity of the dark photon cloud, the spacetime is not that of a Kerr black hole, but gets modified. We, therefore, adopt the solution of a hydrodynamic equilibrium disk in general spacetimes by Abramowicz-Jaroszynski-Sikora (AJS) \cite{abramowicz1978relativistic} as our initial condition, which is parametrized by the specific angular momentum $l_{\rm AJS}$ in the disk, and energy potential $W_{\rm in}$ at inner edge. More details on the initial configuration are given in Appendix \ref{app:disk}.

In this set of simulations, we rescaled the MHD quantities for numerical accuracy, so that the magnetic field is in units of $B_0 = 10^{-8} M^{-1} $ and relative densities are normalized by $\rho_0 = 10^{-16} M^{-2} $. In cgs units, for an $M=10^9 M_{\odot}$ supermassive black hole, $B_0 \approx 235 $ Gauss and $\rho_0 \approx 6 \times 10^{-17} $g/cm$^3 \approx 3.4 \times 10^{10} m_e$/cm$^3$. In the simulations we present, the maximum magnetic field near the jet region is around 0.1 $B_0$ and the maximum density inside the disk is around 0.5 $\rho_0$.

\begin{figure*}
    \centering
    \includegraphics[width = \textwidth]{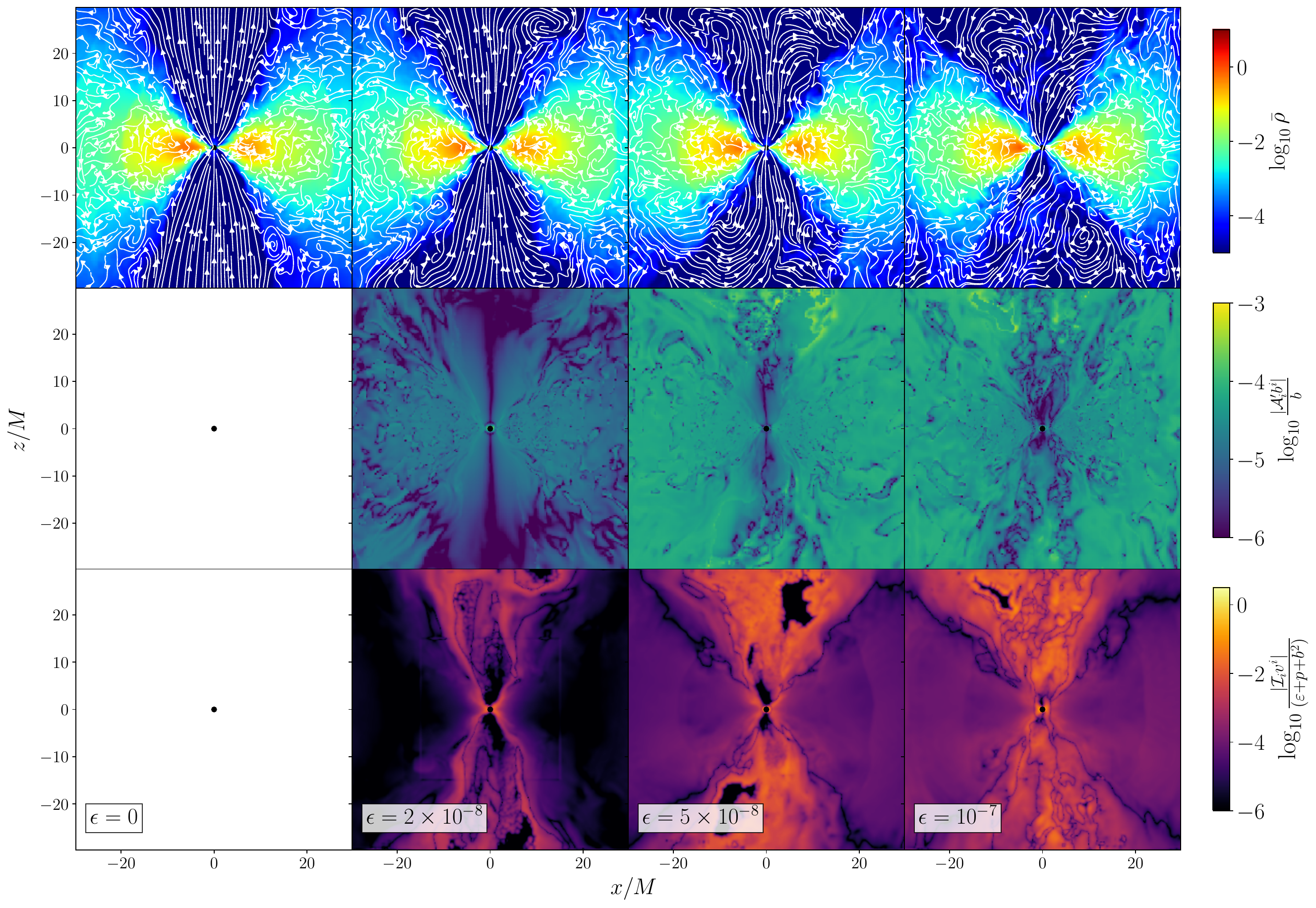}
    \caption{Influence of dark photon on accretion disks at different
      coupling strength $\epsilon = 0, 2\times 10^{-8}, 5\times 10^{-8},
      10^{-7}$ at a late time $t = 6016M$ after the coupling is turned on
      at $t=2070M$, where $M$ is the black hole mass. Top row: fluid density normalized by reference $\rho_0
      = 10^{-16} M$ and visible magnetic field lines $\vec B$ shown in
      white. Middle row: magnitude of fluid magnetization $\sigma =
      b^2/\rho$ and velocity streamlines. Bottom row: dark Ohmic heating rate, $\mathcal{I}^i v_i$ acting on the fluid normalized by $(\varepsilon + p +
      b^2)$. }
    \label{fig:eps_comparison}
\end{figure*}

The system we simulate features a simple advection dominated accretion flow, which does not become magnetically arrested \cite{tchekhovskoy2011efficient}. The accretion flow can roughly be split into a jet and disk region, as can be seen from the magnetization $\sigma$ (see Fig. \ref{fig:sigma_bh_acc}). The background accretion state is shown in the left most panel of Fig. \ref{fig:eps_comparison}. We evolve the system into an quasi-steady accretion state as indicated by a constant mass accretion rate, $\dot{M}_{\rm BH}$. While different accretion states could be considered, these are typically more dynamical and can feature periodic erruptions \cite{tchekhovskoy2011efficient,Ripperda:2021zpn}, which may make an initial analysis of potential dark matter effects harder to quantify.

After enabling the dark photon coupling, we evolve the accreting system into a new quasi-steady state. We point out that for the couplings and mass hierarchies we explore, the dark photon cloud never dominates the accretion flow as it does in the previous Section. Consequently, the effects we probe by construction will largely act as corrections/perturbations to the background flow.\\
We quantify the impact of the dark photon field in different ways. First, we aim to quantify the impact on the standard model magnetic field, as well as on the energy transfer between the plasma and the cloud.

In the accretion disk simulations we study in this work, only the effect of the
dark Lorentz force, controlled by coupling constant $\epsilon$, is turned
on, while the dynamo-like effect in Eq. \ref{eq_E_dark_current}, controlled
by the rescaled coupling $\bar \epsilon$, is effectively not included, i.e. the
simulations are in the ideal limit with perfect conductivity.
This is because the dominant effect in this accretion state will be the
dark Lorentz force.
In line with our previous discussion,
we justify this assumption by a posteriori estimating an effective $\alpha-$term in the induction equation, which is
given by $\kappa b^\mu$ (see e.g. \cite{Bucciantini:2012sm,Most:2023sme})
with $\kappa \sim A_i b^i/b$. This is shown in the center row of Fig.
\ref{fig:eps_comparison}.  We see the effect is weak inside around $r\sim
10M$ where most cloud energy resides. 

\begin{figure}
    \centering
    \includegraphics[width = 8cm]{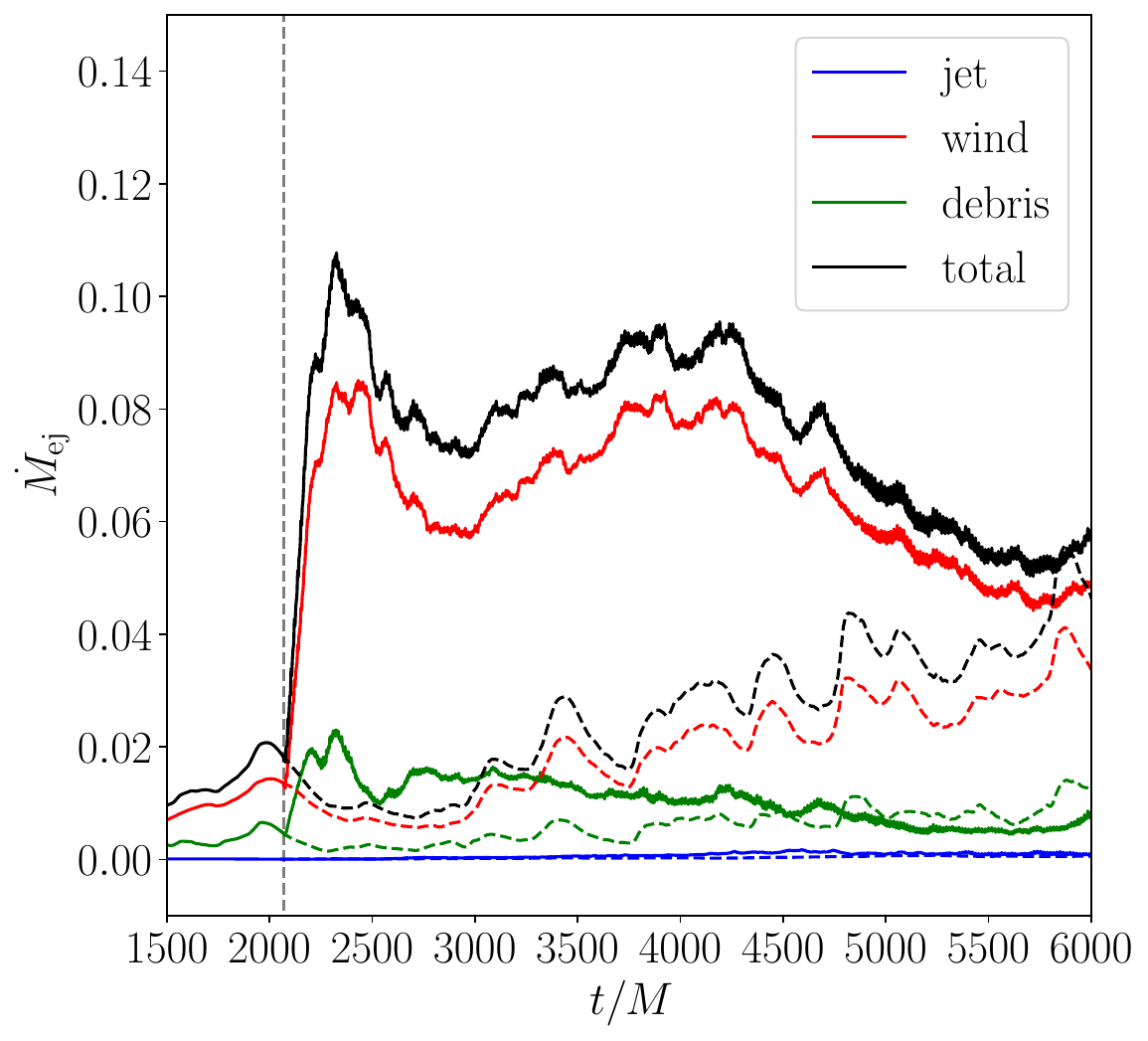}
    \caption{Coupling to the dark photon cloud enhances the outflow rate, $\dot{M}_{\rm ej}$ (top, measured at $r=90M$), while the accretion rate was not significantly changed (bottom, measured at $r=3M$). 
    {Contributions to the outflow: Jet region (blue, integrated $10 ^ \circ $ around z-axis), debris stream (green, integrated $10^\circ$ above and below equatorial plane), and disk wind (red, integrated between the jet and debris boundaries).  } 
    }
    \label{fig:outflow}
\end{figure}

 We can see that the impact of this term scales with the coupling, with stronger couplings providing substantial corrections to the electromotive force. Due to alignment of the magnetic field with the dark photon field, the effect is suppressed in the close vicinity of the black hole, and (where matter-free) also in the jet region. The strongest injection seemingly takes place in low density regions, with the effect being of comparable order throughout the accretion disk, but at least one order lower than in Fig. \ref{fig:FFBH_alpha_sigma}, where the effect was more substantial and dynamically important. \\
In addition to changes to the induction equation, the dark Lorentz force can also act on the accretion flow. We quantify this effect by computing the effective dark heating term $I_i v^i$ (bottom row, Fig. \ref{fig:eps_comparison}).
Looking at different coupling strengths, we can see that the impact on the disk is always subdominant, however in the jet sheath region, as well as in the funnel for higher couplings we see up to $10\%$ energy injection rate relative to the enthalpy, $\varepsilon + p + b^2$ of the fluid. This leads us to speculate that the impact on low density regions is likely strongest in this context.

\begin{figure}
    \centering
    \includegraphics[width = 0.5\textwidth]{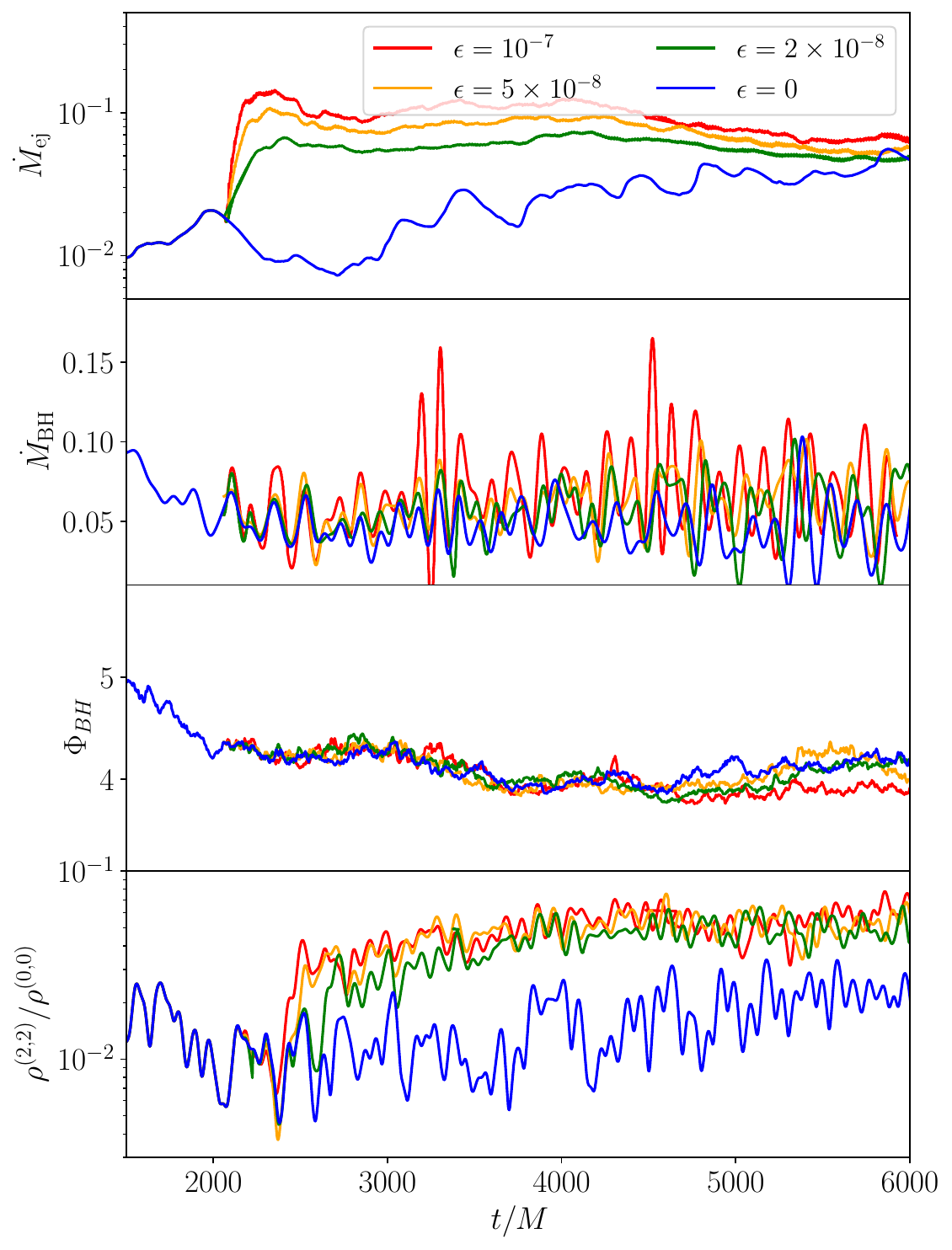}
    \caption{ {Shown are: Mass outflow rate $M_{\rm ej}$ across the surface at $r=90M$ for different coupling constants $\epsilon = $ $10^{-7}$(red), $5\times 10^{-8}$ (orange) $2\times 10^{-8}$ (green), and no coupling for reference (blue), the mass accretion rate, $\dot{M}_{\rm BH}$ and the magnetic flux $\Phi_{\rm BH} = \int |B^r| \sqrt{-\gamma} d\Omega $ across the surface near horizon at $r=3M$, the $(\ell,m)=$(2,2) harmonic oscillation modes of fluid density, $\rho$ extracted at $r=10M$. All times, $t$, are stated relative to the mass $M$ of the black hole. }
    }
    \label{fig:Phi_BH_eps}
\end{figure}

\begin{figure}
    \centering
    \includegraphics[width = 0.5\textwidth]{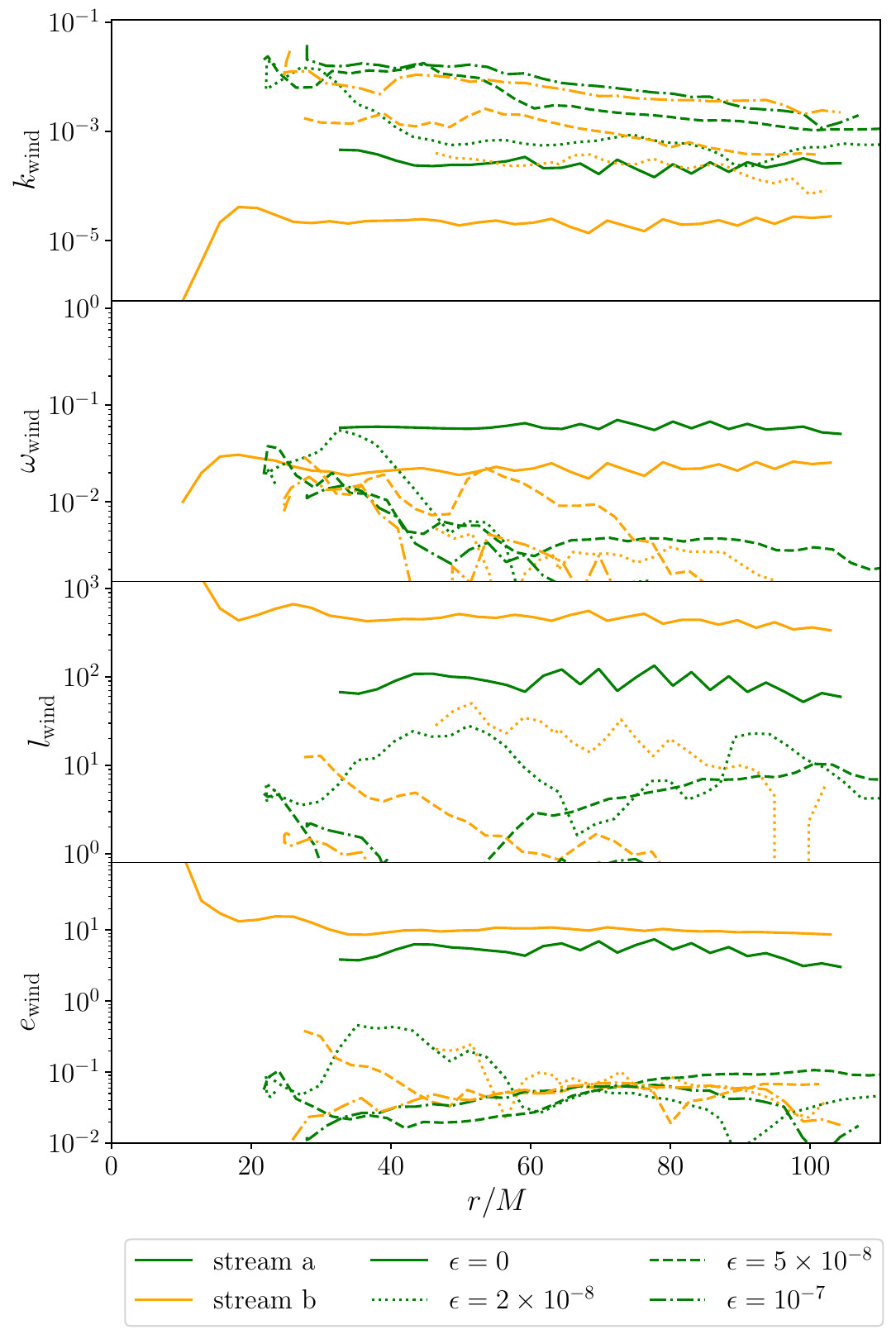}
    \caption{Conserved wind quantities along the the yellow and green streamline shown in Fig. \ref{fig:wind_t5000}. These include the mass loading parameter $k_{\rm wind}$, the effective rotation rate $\omega_{\rm wind}$, the specific angular momentum $l_{\rm wind}$, and the Bernoulli energy density $e_{\rm wind}$. Dashed lines indicate simulations with non-zero dark photon coupling $\epsilon$. We can see that in the presence of a dark matter coupling, especially the rotation quantities, $\omega_{\rm wind}$ and $l_{\rm wind}$, are no longer conserved. }
    \label{fig:wind_trajectories}
\end{figure}


\begin{figure*}[p]
    \centering
    \includegraphics[width=18cm]{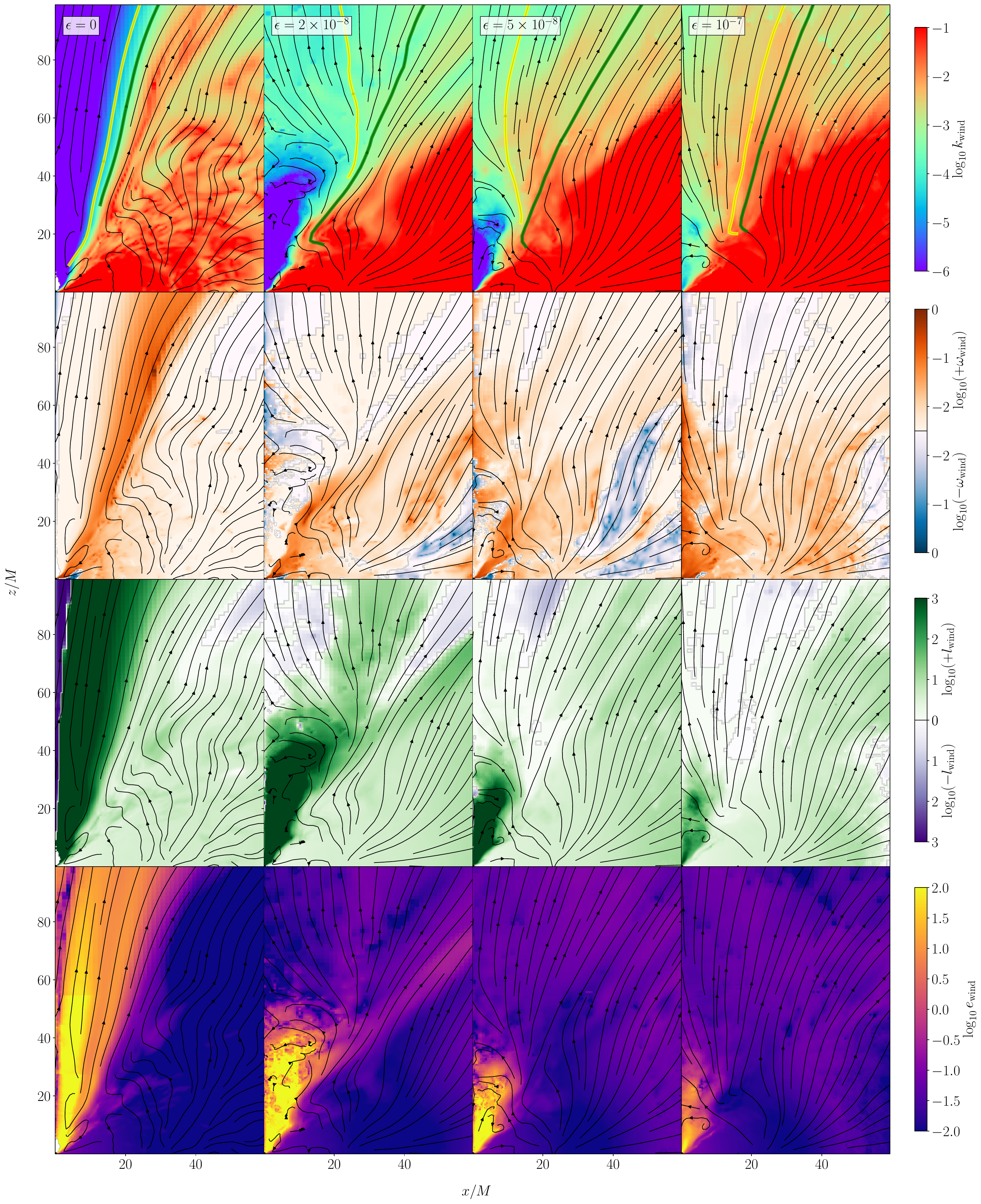}
    \caption{The conserved quantities in the wind region for different coupling strengths, averaged over time $4857.6M < t <  5216.0M$, where $M$ is the black hole mass. The velocity (black) streamlines are shown for reference. In the case with no coupling (first column), $k_{\rm wind}, \omega_{\rm wind},l_{\rm wind},e_{\rm wind}$ are approximately conserved along fluid streams, i.e. velocity streamlines appear as equipotential contours. However, when the coupling is turned on ($\epsilon = 2 \times 10^{-8}$, $5 \times 10^{-8}$, $ 10^{-7} $, in second, third and fourth columns), the wind quantities are no longer conserved along fluid streams. We quantify this in Fig. \ref{fig:wind_trajectories} by tracing the conserved quantities along the streams highlighted by thick yellow/green lines on the top panels.}
    \label{fig:wind_t5000}
\end{figure*}

We now briefly review some global properties of the accretion flow. For the disk properties, we further quantify the influence of the dark sector on the outflow, accretion rate, oscillation modes, and the magnetic flux threading the black hole (see Fig. \ref{fig:Phi_BH_eps}).
We begin with the mass outflow/ejection rate, $\dot{M}_{\rm ej}$, which we measure on a spherical surface at a distance $r=90M$ from the black hole. We can see that the outflow rate is roughly constant for most of the simulation without dark photon coupling. Once the coupling is turned on, especially in the wind region, the outflow is significantly enhanced, increasing by one order of magnitude.
We further quantify where this outflow is sources by looking at projections in various azimuthal bins (see Fig. \ref{fig:outflow}).
We can clearly see that jet and equatorial outflows are unaffected by the dark photon field. Instead, the main effect is driving a wind (likely from the disk edge) where we found the energy injection from the dark cloud to be substantial (see Fig. \ref{fig:eps_comparison}).
Overall, the amount of outflows scales with the coupling, but not very strongly so, with variations in the outflow rate due to changes in $\epsilon$ being less than a factor three.
Also, both Fig. \ref{fig:outflow} and \ref{fig:Phi_BH_eps} show that the outflow saturates after $\sim 500M$ time of coupling with the dark sector, and enters a relatively stable stage where the dynamics are altered by Proca field. 
We can contrast this behaviour with the mass accretion rate on the black hole, $\dot{M}_{\rm BH}$. Consistent with the absence of modifications in the jet region, the mass accretion rate onto the black hole seems entirely unaffected by dark photon interactions. This lets us draw the important conclusion that likely jet luminosities (in the coupling regimes we consider) will not be affected by the dark photon field.
Consistent with this assessment, the magnetic flux on the black hole $\Phi_{\rm BH} =   \int |B^r| \sqrt{-\gamma} d\Omega $ neither decreases nor increases due to the presence of a dark photon coupling. Meaning that regions close to the black hole, where the standard model magnetic field will be strongest, are indeed unaffected.
Finally, we do observe bar shaped disk oscillations driven by the decaying Proca field in the disk, as quantified by the $l=m=2$ mode of the rest-mass density in the disk. However we caution that the oscillation strength seems independent of the coupling, and the change is less than a factor $2$ compared to the no coupling case, making the effect potentially subleading.


{
Having found that the dark Lorentz force alters the wind ejection, we now
attempt a more detailed assessment of hydromagnetically driven wind
properties. To this end, we perform an analysis commonly used for
protoplanetary disks (e.g., \cite{Zhu_2018}).\\

The dark Lorentz force $ \vec{ \mathcal{I} } $ continues to drive the evolution of accretion disks in the wind region at late times. 
When the coupling to the dark photon is absent, dynamics of disk wind can be well-approximated by the analytic solutions of Blandford-Payne process for axisymmetric flows \cite{blandford1982hydromagnetic}. In a steady state, the fluid streams follow magnetic field lines in the wind region and there are four conserved quantities along the field lines\cite{Mestel1961,Mestel1968}
\begin{align}
    k_{\rm wind} & =  \frac{\rho v_p}{B_p}, \label{eq_kwind}\\
    \omega_{\rm wind} & = \frac{v_\phi}{R} - \frac{B_\phi v_p}{B_p R},\label{eq_wwind}\\
    l_{\rm wind} & = R \left( v_\phi - \frac{B_\phi}{k_{\rm wind}} \right),\label{eq_lwind}\\
    e _{\rm wind}& = \frac{v^2}{2} + h + \Phi_g + \frac{B_\phi^2}{\rho} - \frac{B_\phi v_\phi }{k_{\rm wind}},\label{eq_ewind}
\end{align}
where the velocity and magnetic field are decomposed into poloidal and toroidal components $\vec B = \vec B_p + B_\phi \hat \phi$, $\vec v = \vec v_p + v_\phi \hat \phi$ on a cylindrical coordinate $(R,\phi,z)$. 


Using this formalism, we can quantify whether or not additional outflows will be dark matter driven, since the presence of a dark Lorentz force will alter the conserved quantities. The dark sector acts as an external supply of energy and angular momentum, so that the combined system does not satisfy the simple symmetries of the Blandford-Payne process. While in principle, a full derivation of the wind quantities in the presence of a dark photon field could be done (assuming the dark photon cloud is approximately axisymmetric), we forgo such an approach in this initial work, and limit ourselves by showing that including dark matter effects leads to a non-conservation especially of the rotational quantities. 

We assess the validity of this approach, by choosing two streamlines (highlighted in green and yellow colors in Fig. \ref{fig:wind_t5000}).
We then compute the various conserved quantities for cases with and without coupling to the dark photon field, which we show in Fig. \ref{fig:wind_trajectories}. By considering a case without coupling, we establish the baseline degree to which our code can conserve the wind quantities. This is show with solid lines in Fig. \ref{fig:wind_trajectories}. We observe that conservation holds extremely well for all quantities along their respective fluid streamlines. 
In numerical simulations of accretion disks, it is also observed that the four quantities above are conserved along fluid streams in a statistical sense after a time/azimuthal average \cite{Anderson_2005,Zhu_2018}. We have also confirmed this in the fiducial simulation with no coupling to dark sectors.

We now perform the same analysis for similar streamlines in the simulations with active coupling, $\epsilon >0$.
In this case, the standard model Lorentz force gets supplemented with its dark equivalent altering the force balance of the system. 

We observe that with coupling enabled the mass loading parameter $k_{\rm wind}$ increases by at least one order of magnitude. Consistent with the observation that substantial disk winds are present in those simulations. We can also nicely see this by looking at a spatial distribution of $k_{\rm wind}$ in Fig. \ref{fig:wind_t5000}. We can see that with enabled coupling the wind launching increases mass loading every where with clear break outs from the disk visible. This effect increases monotonically with the coupling parameter $\epsilon$.\\
We can further see that the energy density, $e_{\rm wind}$, is well conserved, albeit with a substantial offset compared to the non-coupled simulation indicating that we are consistently missing out on an energy contribution from the dark photon cloud.
Indeed, we already saw in Fig. \ref{fig:eps_comparison}, the energy transfer between the dark and visible sectors is particularly strong on surfaces of the disk. 

The main difference is associated with the enhanced wind launching, namely the change in specific angular momentum $l_{\rm wind}$.

}

The modes of superradiant dark photon around highly spinning black holes
are of high angular momentum, having the largest overlap with spherical
harmonic $\ell = m =1$ mode. In other words, the dark Lorentz force highly
modifies the toroidal part of the force. Therefore, the
angular quantities $\omega_{\rm wind}$, $l_{\rm wind}$, shown in the second
and third rows, are changed by several orders of magnitude. Some streams
even go counter-rotating at large radii.  In Fig. \ref{fig:wind_t5000}, we
see a layer of anti-rotating streams, with opposite sign of angular
$\omega'_{\rm wind}$ and $l'_{\rm wind}$. This is driven by the dark
Lorentz force in the toroidal direction. The dark magnetic field is mostly
in poloidal direction $\vec B' \approx \vec B'_p $. On a mathematical
level, the interaction alters the magnetic field configuration so that the
vector potential has nonzero poloidal component $\vec {\mathcal{A}_p} \neq
0$, thus generating a force in the toroidal direction $\epsilon \mu^2 \vec
{\mathcal{A}}_p \times \vec B'_p $ which alters the angular quantities.
Overall we find a strong reduction of $l_{\rm wind}$ with increasing dark photon coupling strength.

\section{Conclusion}
\label{p_conclusion}

In this work we have numerically investigated the potential impact of a superradiant dark photon cloud on different black hole accretion flows. 
Superradiance, originally proposed from black hole perturbation theory calculations \cite{PhysRevD.7.949} and recently further corroborated by numerical simulations \cite{east2017superradiant,east2017superradiant2,witek2013superradiant}, has been the prominent feature of massive bosonic fields around spinning black holes. As a first series of simulations based on our formulation, we studied the field configuration, energy distribution and decay curves of cloud total energy and vector field radiation during the evolution of dark photon cloud in the presence of astrophysical accretion flows. 
We have done so in several steps.
First, we have derived a formulation of the coupled Einstein, Proca and GRMHD (dark magnetohydrodynamics) systems in the nearly perfectly conducting limit. The resulting system retains a dark Lorentz force and a dynamo-like modification of the induction equation, through which the massive dark photon (Proca) evolution can feedback on the plasma.
This dark magnetohydrodynamics system was then implemented into a full 3+1 numerical relativity code (based on the \texttt{EinsteinToolkit}\cite{loffler2012einstein}, \texttt{Canuda}\cite{Canuda_zenodo_3565474}, and \texttt{FIL} \cite{Most:2019kfe}). 

We have then numerically solved the dark magnetohydrodynamics system for two different
black hole accretion flows in the presence of a dark photon cloud.  First,
we have focused on a case where the dark photon cloud drives accretion in a
nearly force-free plasma, similar to a case considered by Ref.
\cite{siemonsen2023dark}. We found that the dark photon cloud can
substantially drive magnetic field amplification and alter the structure of
the accretion flow via the dark Lorentz force.  In the main part of our
work, we have considered a simple advection dominated accretion flow onto a
supermassive black hole. In this scenario, we have investigate the
potential impact of the dark photon cloud onto the flow. We found that the
dark cloud will mainly act on the flow via the dark Lorentz force, driving
oscillations and outflows from the disk. We also confirmed by means of a
detailed wind analysis following
Refs.\cite{blandford1982hydromagnetic,Zhu_2018}, that especially the
angular velocity of the wind is directly impacted by the dark Lorentz
force. This leads us to speculated that wind outflows and potential X-ray
signatures from the disk surface may be impacted.

Overall, this work is intended a first demonstration of the
general-relativistic dark magnetohydrodynamic framework. As such, we have
limited ourselves to an idealized accretion scenario. On of the key
objectives of using our framework is to provide potential actual signatures
of superradiant dark photon clouds. Potential applications could be studies
of low luminosity accretion flows observable comparable to those observable with the Event Horizon
Telescope \cite{EventHorizonTelescope:2019pgp,eht,EHTcode}. 
These will require studies of a magnetically arrested regime
\cite{tchekhovskoy2011efficient}, which we have not considered here. Other
applications could be stellar mass mergers in AGN disks
\cite{2023ApJ...944L..42L,Dittmann:2023sha}, 
for which potential interactions with the flow and jet launching
from the black hole may be relevant. Such a
scenario (without gaseous accretion flows) has been speculated earlier to
results in additional X-ray signatures \cite{siemonsen2023dark}.  
 Superradiance around binary systems is also shown to be possible
 \cite{ribeiro2022binary}, and may affect circumbinary disk accretion flows
 \cite{Bamber:2022pbs}.

\begin{acknowledgements}
The authors are grateful to W. East, J. Huang and A. Philippov for insightful discussions. Simulations were performed on the Sherlock clusters at Stanford University. SX acknowledges support by US Department of Energy under contract DE–AC02–76SF00515. ERM acknowledges support by the National Science Foundation under grants No. PHY-2309210 and AST-2307394, as well as on the NSF Frontera supercomputer under grant AST21006, and on Delta at the National Center for
Supercomputing Applications (NCSA) through allocation PHY210074 from the
Advanced Cyberinfrastructure Coordination Ecosystem: Services \& Support
(ACCESS) program, which is supported by National Science Foundation grants
\#2138259, \#2138286, \#2138307, \#2137603, and \#2138296.
\end{acknowledgements}

\appendix

\section{3+1 decomposition of fluid equations with external energy-momentum flux}
\label{app_fluid}
Eq. \ref{eq_fluidS}, \ref{eq_fluidSi} follows from similar derivations for 3+1 decomposition of energy-momentum conservation in most textbooks\cite{gourgoulhon2007}. We first list some useful properties that are repeatedly used:

\begin{align}
    \nabla_\mu n_\nu = & - K_{\mu\nu} - n_\mu D_\nu \ln \alpha\,, \\
    \nabla_\mu n^\mu = & -K\,,\\
    n^\mu \nabla_\mu n_\nu = & D_\nu \ln \alpha\,, \\
    \sqrt{\gamma}^{-1} \partial_t \sqrt{\gamma} = & D_j \beta^j - \alpha K\,, \\
    D_j Q^j_i = & \sqrt{\gamma}^{-1} \partial_j ( \sqrt{\gamma} Q^j_i ) - \Gamma^k_{ij} Q^i_k\,, 
\end{align}
with $Q$ an arbitrary (1,1) tensor and $\Gamma^k_{ij} = \frac 12 \gamma^{k\ell} (\partial_i \gamma_{\ell j} + \partial_j \gamma_{i\ell} - \partial_\ell \gamma_{ij} ) $ the Christoffel symbol corresponding to spatial metric.

\subsection{Energy equation}
Projecting the equations on $n^\nu$ and expand $T^\mu_\nu$ we get
\begin{equation}
\begin{aligned}
\label{eq_fluidSraw}
    -\mathcal{I}_\phi =& n^\nu \nabla_\mu T^\mu_\nu\,, \\
    =& n^\nu \nabla_\mu S^\mu_nu + n^\mu n^\nu \nabla_\mu S_\nu - \nabla_\mu S^\mu + KS - n^\mu \nabla_\mu S\,,
\end{aligned}
\end{equation}

Using $S^\mu_\nu n^\nu = 0$ and $n^\nu S_\nu = 0 $, we have
\begin{align}
    n^\nu \nabla_\mu S^\mu_\nu = & - S^\mu_\nu \nabla_\mu n^\nu = K^{\mu\nu} S_{\mu\nu}\,, \\
    n^\mu n^\nu \nabla_\mu S_\nu = & - n^\mu S_\nu \nabla_\mu n^\nu = - S_\nu D^\nu \ln \alpha \,,
\end{align}
and the 4-divergence can be expressed by spatial covariant derivatives
\begin{equation}
    \nabla_\mu S^\mu = D_\mu S^\mu + S^\mu D_\mu \ln \alpha\,.
\end{equation}

Together with Eq. \ref{eq_fluidSraw} we obtain

\begin{equation}
    \partial_t S - \beta^i D_i S + \alpha \left( D_\mu S^\mu - KS - K^{\mu\nu} S_{\mu\nu} - \mathcal{I}_\phi \right) + 2 S^\mu D_\mu \alpha = 0\,.
\end{equation}

To rewrite this into flux-conservative form, we collect
\begin{equation}
    \partial_t S - D_i (\beta^i S - \alpha S^i) + S D_i \beta^i - \alpha KS - \alpha K^{ij} S_{ij} - \alpha \mathcal{I}_\phi  + S^i D_i \alpha =0\,,
\end{equation}
and use $ D_i (\beta^i S - \alpha S^i) = \sqrt{\gamma}^{-1} \partial_i \Big[ \sqrt{\gamma } (\beta^i S - \alpha S^i) \Big] $, $ D_j \beta^j - \alpha K  = \sqrt{\gamma}^{-1} \partial_t \sqrt{\gamma}  $, we finally arrive at

\begin{equation}
\begin{aligned}
    &\partial_t \left( \sqrt\gamma S\right) - \partial_i \left[ \sqrt\gamma (\beta^i S - \alpha S^i) \right]\,,  \\
    &\quad \quad \quad = \sqrt{\gamma} \left[ \alpha K^{ij} S_{ij} - S^i D_i \alpha + \alpha \mathcal{I}_\phi \right]\,.
\end{aligned}
\end{equation}

\subsection{Momentum equations}
We have
\begin{equation}
    \gamma^\nu_\alpha \nabla_\mu S^\mu_\nu - K S_\alpha + \gamma^\nu_\alpha n^\mu \nabla_\mu S_\nu - S^\mu K_{\mu\alpha} + S D_\alpha \ln \alpha = \mathcal{I}_\alpha\,.
\end{equation}

By using
\begin{equation}
\begin{aligned}
    D_\mu S^\mu_\alpha := & \gamma^\rho_\mu \gamma^\mu_\sigma \gamma^\nu_\alpha \nabla_\rho S^\sigma_\nu \\ 
    = & \gamma^\nu_\alpha (g^\rho_\sigma + n^\rho n_\sigma ) \nabla_\rho S^\sigma_\nu \\ 
    = & \gamma^\nu_\alpha \nabla_\mu S^\mu_\nu + \gamma^\nu_\alpha n^\rho n_\sigma \nabla_\rho S^\sigma_\nu \\
    = & \gamma^\nu_\alpha \nabla_\mu S^\mu_\nu - S^\mu_\alpha \nabla_\mu \ln \alpha
\end{aligned}
\end{equation}
and
\begin{equation}
\begin{aligned}
    \gamma^\nu_\alpha n^\mu \nabla_\mu S_\nu = & \alpha^{-1} \gamma^\nu_\alpha (\alpha n)^\mu \nabla_\mu S_\nu  + \alpha^{-1} \gamma^\nu_\alpha S_\nu \nabla_\mu  (\alpha n)^\mu \\
    & \quad - \alpha^{-1} \gamma^\nu_\alpha S_\nu \nabla_\mu  (\alpha n)^\mu \\
    = & \alpha^{-1} \gamma^\nu_\alpha \mathcal{L}_{\alpha n} S_\nu - S_\nu K_\alpha^\nu \\
    = & \alpha^{-1} (\partial_t - \mathcal{L}_\beta) S_\alpha - S_\nu K^\nu_\alpha 
\end{aligned}
\end{equation}
we obtain
\begin{equation}
    (\partial_t - \mathcal{L}_\beta ) S_\alpha + \alpha D_\mu S^\mu_\alpha - \alpha K S_\alpha + S^\mu_\alpha D_\mu \alpha + SD_\alpha \alpha = \alpha\mathcal{I}_\alpha\,.
\end{equation}

To rewrite this into flux divergence form, we expand the Lie-derivative and collect
\begin{equation}
    \partial_t S_j + D_i (\alpha S^i_j - \beta^i S_j) + SD_j \alpha - \alpha KS_j + S_j D_i \beta^i - S_i D_j \beta^i = \alpha \mathcal{I}_j\,.
\end{equation}
We can further use that $ D_i (\alpha S^i_j - \beta^i S_j) = \sqrt{\gamma}^{-1} \partial_i \Big[ \sqrt{\gamma} (\alpha S^i_j - \beta^i S_j) \Big] - \Gamma^k_{ji}(\alpha S^i_j - \beta^i S_j)$, and then finally arrive at

\begin{equation}
\begin{aligned}
    & \partial_t  \left( \sqrt\gamma S_j \right) + \partial_i \left[ \sqrt\gamma (\alpha S^i_j - \beta^i S_j) \right] = \\
    & \quad \quad \quad \sqrt{\gamma } \left[ S_i D_j \beta^i - SD_j \alpha + \alpha \mathcal{I}_j +\Gamma^k_{ji}(\alpha S^i_j - \beta^i S_j)  \right]\,.
\end{aligned}
\end{equation}

This can be further simplified (removing the Christoffel symbol) by noting that

\begin{equation}
    D_j \beta^i = \partial_j \beta^i + \Gamma^i_{jk} \beta^k\,,
\end{equation}
and ($S^{i\ell} = S^{\ell i}$ is symmetric)
\begin{align}
    S^i_k \Gamma^k_{ji} = & \frac 12 S^i_k \gamma^{k\ell} \left( \partial_i \gamma_{\ell j} + \partial_j \gamma_{i\ell} - \partial_\ell \gamma_{ij} \right)\,,\\
     = & \frac 12 S^{i\ell} \partial_i \gamma_{\ell j}  + \frac 12 S^{i\ell} \partial_j \gamma_{i \ell }  - \frac 12 S^{i\ell} \partial_\ell \gamma_{i j} \,,\\
     = & \frac 12 S^{i\ell} \partial_j \gamma_{i \ell }\,.
\end{align}
With the right-hand side simplified we get Eq. \ref{eq_fluidSi}

\begin{equation}
\begin{aligned}
    \partial_t  \left( \sqrt\gamma S_j \right) & + \partial_i \left[ \sqrt\gamma (\alpha S^i_j - \beta^i S_j) \right]  \\
    & = \sqrt\gamma \left[ S_i \partial_j \beta^i - S\partial_j \alpha + \frac 12 S^{ik}\partial_j \gamma_{ik} + \alpha \mathcal I_j \right] \, .
\end{aligned}
\end{equation}

\section{Primitive recovery scheme of coupled MHD-Proca system}
\label{p_c2p}

One crucial aspect of any relativistic MHD algorithm is the computation of primitive quantities, such as $\rho$, $u^\mu$ and $h$ from the evolved variables. This involves numerical non-linear root finding, which has been well studied in the relativistic case \cite{Noble:2005gf,palenzuela2015effects,ripperda2019general,kastaun2021robust,ng2023hybrid}.
In the following, we extend such a recent scheme \cite{kastaun2021robust} to the dark MHD case.

The MHD energy momentum tensor in terms of electric and magnetic field, is \cite{Shapiro2003}
\begin{equation}
\begin{aligned}
    T^{\mu\nu}_{\rm MHD} =& \rho h u^\mu u^\nu +  \frac 12 (E^2 + B^2 + 2 p) g^{\mu\nu} \\ 
    &- E^\mu E^\nu - B^\mu B^\nu  + ( n^\mu \varepsilon^{\nu \alpha \beta } + n^\nu \varepsilon^{\mu \alpha \beta } ) E_\alpha B_\beta
\end{aligned}
\end{equation}

The dark current prescription Eq.\ref{eq_ideal}, transformed to the normal frame, means the usual ideal MHD prescription $ E^i = - \varepsilon_{ijk} v_i B_k$ is replaced by
\begin{equation}
\label{eq_Eic2p}
    E^i = - \varepsilon_{ijk} v_i B_k + \bar \epsilon \mu^2 \mathcal{A'}^i \Gamma^{-1}\,,
\end{equation}
where $\bar \epsilon = \epsilon/\sigma$ and remember the spatial part of Proca field $\mathcal{A'}^\mu = A'^\mu + n^\mu n^\nu A'_\nu $.

The evolved variables (conservatives) are

\begin{equation}
    \rho^* = \rho \Gamma \,,
\end{equation}
\begin{equation}
    \tau = S - \rho^* = \rho^* (h\Gamma -1) - p + \frac{E^2+B^2}{2}\,,
\end{equation}
\begin{equation}
    S_i = \rho^* h \Gamma v_i + \varepsilon_{ijk} E^j B^k\,,
\end{equation}

To solve the primitives ($\rho,p,h,v^i$), we follow the idea of \cite{ripperda2019general,kastaun2021robust} by reducing the problem to finding 1D fixed point of a master function. 

We first introduce several shorthand notations: the pure fluid contribution of conservatives
\begin{align}
    \tau_{\rm fluid} = & \rho^*(h\Gamma -1) - p ,\\
    S^{\rm fluid}_i = & \rho^* h \Gamma v_i,
\end{align}
tilded variables rescaled by $\rho^*$
\begin{align}
    \tilde \tau = \frac{\tau}{\rho^*}, \quad \tilde \tau_{\rm fluid} = \frac{\tau_{\rm fluid}}{\rho^*},\quad  \tilde E_i = \frac{E_i}{\sqrt{\rho^*}},
\end{align}
\begin{align}
    \tilde S_i = \frac{S_i}{\rho^*},\quad \tilde S^{\rm fluid}_i = \frac{S^{\rm fluid}_i}{\rho^*} , \quad  \tilde B_i = \frac{B_i}{\sqrt{\rho^*}},
\end{align}
shorthand for rescaled Proca field
\begin{equation}
    \tilde a^i = \frac{\bar \epsilon \mu^2 \mathcal{A'}^i}{\rho^*},
\end{equation}
and since $\bar\epsilon$ here is a small number we ignore terms higher than $O(\tilde a^2)$.

The 1D fixed point strategy we use is introducing a variable $\mu$, compute estimators of primitives $\hat v^i, \hat h , ...$ as if $\mu = 1/(h\Gamma)$ and finally compute the estimator of $\hat \mu$ itself. To get the fixed point of $\mu$ we find the root of a master function $f(\mu) = \mu - \hat\mu$. The scheme we introduce below will smoothly reduce to Kastaun's \cite{kastaun2021robust} when $\tilde a \rightarrow 0$.

We first note two relations: the relation between the triple products $ [\tilde S \tilde a \tilde B] := \epsilon_{ijk} \tilde S^i \tilde a^j \tilde B^k $ and $ [\tilde v \tilde a \tilde B] := \epsilon_{ijk} [ v^i \tilde a^j \tilde B^k] $:
\begin{equation}
    [\tilde S \tilde a \tilde B] = \frac{1+\mu \tilde B^2}{\mu} [v \tilde a \tilde B] + O(\tilde a^2), 
\end{equation}
and the relation between components of $\tilde S^i$ and $v^i$ parallel to $\tilde B$ (define the parallel component as $\tilde S^i _{\parallel} := \frac{\tilde S^j \tilde B_j}{\tilde B^2} \tilde B^i$ ):
\begin{equation}
    \tilde S^i_{\parallel} = \frac{\tilde B^j v_j}{\mu \tilde B^2} \tilde B^i, \quad {\rm therefore}\quad \tilde S^2_\parallel = \frac{ (\tilde B^i v_i)^2 }{\mu^2 \tilde B^2}. 
\end{equation}

Then expand the perpendicular component of $\tilde S_i$, $\tilde S^i_{\perp} := \tilde S^i - \tilde S^i_{\parallel}$, and replace the $\tilde B^i v_i$, $[v \tilde a \tilde B]$ by the two relations above. This gives a quadratic equation for the Lorentz factor $\Gamma^{-1}$ (note $v^2 = 1-\Gamma^{-2}$, and we use a short hand $x = \frac{1}{1+\mu \tilde B^2}$ )
\begin{align}
    \tilde S^2_{\perp} = \frac{1}{x^2 \mu^2} (1-\Gamma^{-2}) - \frac{\tilde S^2_{\parallel}}{x^2} - [\tilde S \tilde a \tilde B] \Gamma^{-1},
\end{align}
with discriminant
\begin{equation}
    \Delta = 1- x^2 \mu^2 \tilde S_\perp ^2 - \mu^2 \tilde S^2_\parallel + \frac{x^4 \mu^4}{4} [\tilde S \tilde a \tilde B]^2.
\end{equation}
Our estimator for Lorentz factor is
\begin{equation}
    \hat \Gamma^{-1} = \sqrt{\hat \Delta} - \frac{x^2\mu^2[\tilde S \tilde a \tilde B]}{2},
\end{equation}
compared with \cite{kastaun2021robust}, setting an upper limit for the velocity $v^2<v_0^2 := \frac{\tilde S^2}{h_0^2 + \tilde S^2} $ (with $h_0$ the lower bound of relativistic enthalpy) is equivalent to setting a lower limit for the discriminant

\begin{equation}
    \hat \Delta = {\rm max} \{\,\Delta\,, \left(\sqrt{1-v_0^2}+\frac{x^2\mu^2[\tilde S \tilde a \tilde B]}{2}\right)^2 \}\,,
\end{equation}
which also guarantees the safety of taking square root of $\hat \Delta$ for arbitrary $\mu$. Thus the estimator for velocity and fluid momenta will be
\begin{align}
    \hat v^2 = 1-\hat \Gamma^{-2} , \quad \tilde S^2_{\rm fluid} = \frac{\hat v^2}{\mu^2}.
\end{align}
The $1/\mu^2$ in $\tilde S^2_{\rm fluid}$ may impair numerical safety when $\mu\rightarrow 0$, but the $1/\mu^2$ actually cancel once we expanding everything out, i.e. in code implementation we actually use
\begin{align}
    \tilde S^2_{\rm fluid} = & x^2 \tilde S^2_\perp + \tilde S^2_\parallel + x^2 [\tilde S \tilde a \tilde B] \sqrt{\hat \Delta }, \\
    \hat v^2 = & {\rm min} \{ \mu^2 \tilde S^2_{\rm fluid} , v_0^2 \}\,,
\end{align}
The equation above may still be negative on rare occasions when the assumed small quantity $[\tilde S \tilde a \tilde B]$ became large negative numbers compared to $\tilde S^2_\perp $. In that case we floor $\tilde S^2_{\rm fluid}$ by 0.

Expanding the electric field Eq. \ref{eq_Eic2p}
\begin{equation}
\begin{aligned}
    \tilde E^2 =& \tilde B^2 v^2 - (\tilde B \cdot v)^2 + 2 \Gamma^{-1} [v\tilde a \tilde B] + O(\tilde a^2)\,, \\
    =& \tilde B^2 \hat v^2 - \mu^2\tilde B^2 \tilde S^2_{\parallel} + 2 \hat \Gamma^{-1} \frac{\mu \, [\tilde S \tilde a \tilde B]}{1+\mu \tilde B^2}\,,
\end{aligned}
\end{equation}
we also have the estimator for fluid energy
\begin{equation}
    \tilde \tau_{\rm fluid} =  \tilde \tau - \frac{\tilde E^2 + \tilde B^2}{2}\,,
\end{equation}

Then we have the estimator for the density $\hat \rho$, internal energy $\hat \epsilon$
\begin{align}
    \hat \rho = & \rho^* \hat \Gamma^{-1}\,, \\
    \hat \epsilon = & \hat \Gamma (\tilde \tau_{\rm fluid} - \mu \tilde S^2_{\rm fluid}) + \frac{\hat v^2 \hat \Gamma^2}{1+\hat \Gamma }\,,
\end{align}
and pressure $\hat p$, specific enthalpy $\hat h$ through equation of state
\begin{align}
    \hat p = & p(\hat \rho, \hat \epsilon)\,, \\
    \hat h = & 1+\hat \epsilon + \hat p / \hat \rho\,.
\end{align}
Finally we re-express $\hat \mu$ itself by

\begin{equation}
    \hat \mu = \frac{1}{\hat h \hat \Gamma^{-1} + \tilde S^2_{\rm fluid} \mu }\,.
\end{equation}

We see that the scheme defined above smoothly reduces to Kastaun's algorithm \cite{kastaun2021robust} when $\tilde a \rightarrow 0$ thus the existence and uniqueness of solution is guaranteed when Proca field is small enough. 

In accordance with Kastaun's algorithm, we make a minor modification to the estimator of $\nu = h \Gamma^{-1}$, from the estimators above, we have
\begin{equation}
    \hat \nu_A = (1+\hat \epsilon +\hat p/\hat \rho )\hat \Gamma^{-1}.
\end{equation}
Noting another estimator of internal energy
\begin{align}
    \hat \epsilon_B = & \hat \Gamma (\tilde \tau_{\rm fluid} - \mu \tilde S^2_{\rm fluid}) + \hat \Gamma -1,
\end{align}
we can multiply the $\nu$ estimator by $\frac{1+\hat \epsilon_B}{1+\hat \epsilon}$ and have
\begin{equation}
    \hat \nu_B = \left( 1+ \frac{\hat p}{\hat \rho (1+\hat \epsilon)} \right) \left( 1+ \tilde \tau_{\rm fluid} - \mu \hat S^2_{\rm fluid} \right) .
\end{equation}
So the actual master function we use is
\begin{align}
    f(\mu) = &\, \mu - \frac{1}{\hat \nu + \mu \tilde S^2_{\rm fluid}}\\
    \hat \nu = &\, {\rm max} \{\hat \nu_A, \hat \nu_B\}
\end{align}

\section{Accretion disk setup}\label{app:disk}

{To set up the accretion disk according to the AJS solution, we first note that the black hole-Proca system can be approximated by a stationary, axisymmetric spacetime
\begin{equation}
    ds^2 = g_{tt} dt^2 + 2g_{t\phi} dtd\phi + g_{\phi\phi} d\phi ^2 + g_{rr} dr^2 + g_{\theta \theta} d\theta^2,
\end{equation}
where the metric components are only dependent on $(r,\theta)$. Assuming an ideal fluid circulating around the black hole only in $\phi$ direction, we have the energy-momentum tensor $T^{\mu\nu} = \rho h u^{\mu } u^\nu - p g^{\mu\nu} $ where the fluid velocity $u^\mu$ only have nonzero $t$ and $\phi$ components. Once the specific angular momentum $l_{\rm AJS}$ is specified, the fluid velocity distribution is fixed by geodesic equations, with energy
\begin{equation}
    \varepsilon_{\rm AJS} = u_t = - \sqrt{ \frac{g_{t\phi}^2 - g_{tt} g_{\phi\phi} }{ g_{\phi\phi} + 2 l_{\rm AJS} g_{t\phi} + l^2_{\rm AJS} g_{tt} }},
\end{equation}
and angular velocity
\begin{equation}
    \omega_{\rm AJS} = \frac{u^\phi}{u^t} = - \frac{g_{tt} l _{\rm AJS} + g_{t\phi} }{ g_{t\phi} l_{\rm AJS} +  g_{\phi \phi} }.
\end{equation}
Then the fluid described by the relativistic Euler equation, as discussed in \cite{abramowicz1978relativistic,kozlowski1978analytic}, is governed by the potential $W_{\rm AJS} (r,\theta) = \ln \varepsilon_{\rm AJS} $. Once the inner edge of the disk, and thereby the maximum potential $W_{\rm in}$, is specified, we set fluid quantities in the places with $W_{\rm AJS} (r,\theta ) > W_{\rm in} $ to atmospheric value. Inside the disk $ W_{\rm AJS} (r,\theta ) < W_{\rm in} $, we set density $\rho$, pressure $p$ and internal energy $\varepsilon$ of the fluid according to AJS solution and a simple polytropic equation of state. In the case without Proca field, we tested the analytic solution in Kerr spacetime as discussed in \cite{kozlowski1978analytic}. When the spacetime is modified by Proca field, we set a closed accretion disk with parameter $l_{\rm AJS} = 3.22, W_{\rm in} = -0.014141 $, evolved for $2000M$ time before we turn on the coupling to the dark sector. The seed magnetic field is initialized as a poloidal field \cite{2015CQGra..32q5009E}
\begin{equation}
\begin{aligned}
    A_x =& -y\, A_b \,\left[{\rm{max}} (p-p_{\rm cut},0)\right]^{n_s},\\
    A_y =&\,\, x\, A_b \, \left[{\rm{max}} (p-p_{\rm cut},0)\right]^{n_s},\\
    A_z =&\,\, 0,\\
\end{aligned}
\end{equation}
where the pressure $p$ is the fluid pressure, and we use $n_s=2$, $p_{\rm
cut}= 0.04 \max p$ and $A_b$ such that maximum magnetization $\bar \sigma
\approx 50$. 
This allows us to fully resolve the magneto-rotational instability (MRI) over a long-time ($>6000M$) evolution.

\bibliographystyle{unsrt}
\bibliography{refs}

\end{document}